\documentclass{aa}
\usepackage{txfonts}                    

\usepackage{natbib}
\usepackage{hyperref}        
\usepackage{multirow, makecell}    

\usepackage{graphicx}

\defcitealias{1983SoPh...88..179E}{ER83}
\defcitealias{2002ApJ...577..475R}{RR02}
\defcitealias{2008ApJ...676..717L}{L08}
\defcitealias{2008ApJ...679.1611T}{T08}
\defcitealias{2015A&A...582A.120S}{SL15}
\defcitealias{2023MNRAS.518L..57L}{paper I}

\newcommand{\Exp}[1]{\ensuremath{{\rm e}^{#1}}}
\newcommand{\mathd}{\ensuremath{{\rm d}}}

\newcommand{\uvec}[1]{\vec{e}_{#1}}

\newcommand{\Alf}{Alfv$\acute{\rm e}$n}

\newcommand{\Ampere}{Amp$\acute{\rm e}$re}

\newcommand{\kappae}{\ensuremath{\kappa_{\rm e}}}
\newcommand{\kappai}{\ensuremath{\kappa_{\rm i}}}

\newcommand{\kUniMedia}{\ensuremath{\mathsf{k}}}

\newcommand{\mi}{\ensuremath{m_{\rm i}}}
\newcommand{\me}{\ensuremath{m_{\rm e}}}

\newcommand{\myni}{\ensuremath{n_{\rm i}}}

\newcommand{\omgtwoD}{\ensuremath{\omega^{\rm 2D}}}
\newcommand{\omghat}{\ensuremath{\hat{\omega}}}

\newcommand{\metwoD}{\ensuremath{m_{\rm e}^{\rm 2D}}}

\newcommand{\rhoi}{\ensuremath{\rho_{\rm i}}}
\newcommand{\rhoe}{\ensuremath{\rho_{\rm e}}}

\newcommand{\vecvph}{\ensuremath{\vec{v}_{\rm ph}}}
\newcommand{\vecvgr}{\ensuremath{\vec{v}_{\rm gr}}}

\newcommand{\vgy}{\ensuremath{v_{{\rm gr},y}}}
\newcommand{\vgz}{\ensuremath{v_{{\rm gr},z}}}

\newcommand{\vph}{\ensuremath{v_{\rm ph}}}
\newcommand{\vA }{\ensuremath{v_{\rm A}}}
\newcommand{\vAe}{\ensuremath{v_{\rm Ae}}}
\newcommand{\vAi}{\ensuremath{v_{\rm Ai}}}

\newcommand{\vecfPG}{\ensuremath{\vec{f}^{\rm PG}}}
\newcommand{\fPGx}{\ensuremath{{f}^{\rm PG}_{x}}}
\newcommand{\fPGy}{\ensuremath{{f}^{\rm PG}_{y}}}
\newcommand{\vecfT}{\ensuremath{\vec{f}^{\rm T}}}
\newcommand{\fTx}{\ensuremath{{f}^{\rm T}_{x}}}
\newcommand{\fTy}{\ensuremath{{f}^{\rm T}_{y}}}

\newcommand{\Lambdai}{\ensuremath{\Lambda_{\rm i}}}

\begin{document}

\title{Three-dimensional kink modes in solar coronal slabs: 
       Group velocities and their implications for impulsively excited waves}
\author{          Jing Liu            \inst{1}
            \and  Bo Li               \inst{1}
            \and  Mijie Shi           \inst{1}
            \and  Mingzhe Guo         \inst{2} 
            \and  Hui Yu               \inst{1}
       }

\institute{
    Shandong Key Laboratory of Space Environment and Exploration Technology,
    Institute of Space Sciences, 
    Shandong University, Weihai 264209, China
   \email{bbl@sdu.edu.cn}
\and 
   Center for Integrated Research on Space Science, Astronomy, and Physics, 
   Institute of Frontier and Interdisciplinary Science, Shandong University, Qingdao 266237, China \\
}

\titlerunning{3D kink modes in solar coronal slabs}
\authorrunning{Liu, J. et al.}

\date{Received ......... / Accepted .........}

\abstract
{Little attention has been paid to group velocities of three-dimensional (3D) MHD waves in solar coronal seismology.
}
{This study aims to present a rather comprehensive examination on the group velocities of trapped 3D kink modes in coronal slabs, emphasizing the connection of mode analysis to both mode characterization and impulsively excited 3D kink waves.}
{We work in linear, ideal, pressureless MHD, and take the equilibrium slab to be symmetrically structured only in one transverse direction. The dispersion relation is numerically solved, with the results understood by making in-depth analytical progress. We address both the transverse fundamental and its first overtone.  
}
{We develop a three-subgroup scheme for categorizing 3D kink modes on the plane spanned by the axial and out-of-plane wavenumbers. The group ($\vecvgr$) and phase velocities ($\vecvph$) sit on the same side of the equilibrium magnetic field ($\vec{B}_0$) for  {the ``$\vec{B}_0$-same-side A'' and ``$\vec{B}_0$-same-side F'' subgroups}, which are further discriminated by the directional similarity of $\vecvgr$ and $\vec{B}_0$.  {The ``$\vec{B}_0$-straddling'' subgroup} 
is peculiar in that $\vecvgr$ and $\vecvph$ lie astride $\vec{B}_0$, a feature that cannot be found for waves in unbounded uniform media in pressureless MHD.  {This ``$\vec{B}_0$-straddling'' subgroup} pertains to both the fundamental and its overtones. We further place our results in the context of impulsive waves, employing the method of stationary phase to predict the large-time wavefront morphology in the plane of symmetry of the equilibrium slab. Wavefronts directed toward $\vec{B}_0$ derive exclusively from  {``$\vec{B}_0$-straddling'' modes}, and are confined to narrow sectors.
}
{The directional information of impulsively excited 3D wavefronts carries rich seismic information, whose inversion requires a through understanding of the behavior of group velocities of 3D modes.}


\maketitle 
\nolinenumbers
\section{Introduction}
\label{sec_intro}

The past two decades have witnessed some rapid accumulation 
    of observations on low-frequency waves and oscillations
    in the solar corona
    \citep[see e.g.,][for reviews]{2007SoPh..246....3B,2012RSPTA.370.3193D,2022SoPh..297...20S}.
These observations have largely been placed in two contexts,
    one being the problem of coronal heating 
    \citep[see the reviews by e.g.,][]{2012RSPTA.370.3217P,2015RSPTA.37340261A,2020SSRv..216..140V}, the other being solar coronal seismology
    \citep[SCS; see e.g.,][for reviews]{2020ARA&A..58..441N,2024RvMPP...8...19N}.
Whichever the application, theoretical investigations into linear
    magnetohydrodynamic (MHD) waves in structured media
    often prove indispensable, with the transverse structuring being 
    of particular interest given the filamentary nature
    of the observed wave hosts 
    \citep[see e.g.,][for topical issues]{2011SSRv..158..167E,2022SSRv..218...13N,2023SoPh..298...40K}.

We focus on how the transverse structuring influences 
    wave dispersion, by which we specifically mean
    the distinction between
    the phase and group velocities.
This distinction has long been of theoretical interest in SCS,
    and tends to be largely addressed for the response of
    some density-enhanced equilibria to impulsive localized exciters
    \citep[see e.g., the reviews by][]{2020ARA&A..58..441N,2020SSRv..216..136L}.
Consider one-dimensional (1D) equilibria, 
    for which the density structuring is restricted to only
    one transverse direction.
Let ``2D motions'' refer to those that prohibit wave propagation 
    in the extra transverse direction. 
It has been customary to examine 2D fast sausage motions
    to bring out the dispersive effects, be the equilibria
    of slab or cylindrical geometry
    \citep[e.g.,][]{1983Natur.305..688R, 1984ApJ...279..857R, 1988A&A...192..343E}.
Impulsively excited sausage motions were predicted to possess a     
    tell-tale three-phase signature in their time sequences sampled 
    sufficiently far from the exciters \citep{1983Natur.305..688R}. 
An almost monochromatic ``periodic phase'' appears first, followed
    by a stronger ``quasi-periodic phase'' and eventually by
    some monochromatic ``decay phase''
    \citep[see][for theoretical reasons]{1986NASCP2449..347E}.
This three-phase behavior turns out to be compatible with
    2D time-dependent simulations~\citep[e.g.,][]{1993SoPh..144..101M,1994SoPh..151..305M}. 
Likewise, the predicted periodicities were compatible with
    the observed pulsating behavior in, say, 
    radio bursts~\citep[e.g.,][]{1990SoPh..130..151Z,1990SoPh..130..161F}
    and visible forbidden line intensities \citep[e.g.,][]{1984SoPh...90..325P,1987SoPh..109..365P}.
Seismology was therefore enabled, yielding such key coronal parameters 
    as the transverse \Alf\ time \citep[e.g.,][]{1984ApJ...279..857R,1990SoPh..130..161F}.
    
Considerable progress has been made for the dispersive evolution
    of impulsively excited waves in the new century.
For instance, \citet{2004MNRAS.349..705N} showed that the
    three-phase signature in time sequences
    translates into ``crazy tadpoles'' in the associated Morlet spectra.
Some subtleties notwithstanding, the expected temporal or Morlet features
    were shown to largely persist regardless of the details of 
    the initial exciter \citep[e.g.,][]{2019A&A...624L...4G,2021MNRAS.505.3505K}
    or the wave host \citep[e.g.,][]{2005SSRv..121..115N,2010ITPS...38.2243J,2013A&A...560A..97P,2016ApJ...833...51Y,2017ApJ...836....1Y,2022MNRAS.515.4055G,2025ApJ...990....1S,2026ApJ...996...72S}.
That these features tend to be robust is not surprising, given 
    the robust applicability to large-time signals of 
    the method of stationary phase 
    \citep[MSP;][]{1986NASCP2449..347E,2023MNRAS.518L..57L}.
The scenario of impulsive waves
    has therefore been broadly invoked for or implicated in
    understanding the oscillatory
    behavior observed in, say, 
    the radio~\citep[e.g.,][]{2009ApJ...697L.108M, 2018ApJ...855L..29K}
    and visible passbands~\citep[e.g.,][]{2001MNRAS.326..428W,2003A&A...406..709K,2016SoPh..291..155S}.
Likewise, instances in line with this scenario have been found
    in imaging data, with two examples being the cyclic transverse displacements
    of streamer stalks 
    \citep[streamer waves; e.g.,][]{2010ApJ...714..644C,2013ApJ...766...55K,2020ApJ...893...78D},
    and the quasi-periodic fast propagating waves 
    \citep[QFPs;][]{2010ApJ...723L..53L,2011ApJ...736L..13L,2012ApJ...753...53S}.

This study is intended to offer an in-depth mode analysis of
    3D kink motions in coronal slabs.
Of particular interest are two questions, namely
    ``how is wave dispersion influenced by the inclusion of the third dimension?'' and 
    ``what spatio-temporal features can be expected for impulsive 3D kink motions?''.
 {To our knowledge, the most relevant study is the one by 
    \citet[][hereafter \citetalias{2023MNRAS.518L..57L}]{2023MNRAS.518L..57L},
        where we focused on demonstrating the peculiar propagation features  
        of impulsively excited 3D kink motions via a representative time-dependent simulation.
    This manuscript is distinct from \citetalias{2023MNRAS.518L..57L}
       in both scope and objective.   
    Firstly, mode analysis was performed in \citetalias{2023MNRAS.518L..57L},
        the purpose being primarily to support the interpretation of the specific
        time-dependent simulation results. 
    Accordingly, the mode analysis was conducted largely in a numerical manner
        and was restricted to the transverse fundamental. 
    In contrast, this manuscript is dedicated to a systematic investigation
        into the group velocity behavior of 3D kink modes trapped in a slab configuration.
    Numerical solutions to the pertinent dispersion relation will be presented,
        and we will make substantial analytical progress in multiple asymptotic regimes.
    Furthermore, both the transverse fundamental and its overtones will be addressed.
    We note that transverse overtones have received little attention so far,
        despite the long-lasting interest in oblique kink modes
        \citep[e.g.,][]{1978ApJ...226..650I,1979ApJ...227..319W,1988JGR....93.5423H}.
    Secondly, we will capitalize on our mode analysis to come up with a scheme for categorizing
        3D kink modes from the group velocity perspective,
        thereby complementing their restoring-force-based characterization~\citep[e.g.,][]{2009A&A...503..213G,2017SoPh..292..192B}.
    Thirdly, our theoretical results will be connected to the dispersive evolution of 
        impulsively excited 3D kink motions.
    We will specifically examine what morphological features are expected before
        performing a time-dependent simulation, in contrast to \citetalias{2023MNRAS.518L..57L} where the simulation results were interpreted largely a posteriori.  
}

This manuscript is structured as follows. 
Section~\ref{sec_prob} formulates our problem,
    presenting the relevant dispersion relations (DRs)
    and collecting some necessary definitions.
Section~\ref{sec_results} then details the behavior of
    the group velocities of 3D kink modes by numerically solving
    the DR.
Also presented are a set of approximate analytical solutions,
    which prove valuable for understanding the numerical results. 
Section~\ref{sec_impulsive} places our findings in the context of
    the large-time features of impulsively excited kink motions.
Our study is summarized in Sect.~\ref{sec_conc}, where some concluding remarks
    are also offered.

\section{Problem formulation}
\label{sec_prob}

\subsection{General formulation}
\label{sec_prob_sub_general}
We adopt ideal, gravity-free, pressureless MHD throughout, 
    in which the primitive quantities are the mass density $\rho$,
    velocity $\vec{v}$, and magnetic field $\vec{B}$.
Let the subscript $0$ denote equilibrium quantities.
We consider only static equilibria ($\vec{v}_0 = 0$). 
Let $(x, y, z)$ be a Cartesian coordinate system. 
The equilibrium magnetic field is taken to be uniform and $z$-directed
   ($\vec{B}_0 = B_0 \uvec{z}$).          
We further assume the equilibrium density $\rho_0$ to be a function 
    of $x$ only.
The \Alf\ speed is defined by $\vA^2 = B_0^2/(\mu_0 \rho_0)$,
    with $\mu_0$ being
    the magnetic permeability of free space.

Let the subscript $1$ denote small-amplitude perturbations. 
The linearized, time-dependent, ideal MHD equations write
\begin{eqnarray}
    \rho_0 \dfrac{\partial \vec{v}_{1}}{\partial t}
&=& \dfrac{(\nabla\times\vec{B}_1)\times \vec{B}_0}{\mu_0}
 =  -\dfrac{\nabla(\vec{B}_0\cdot\vec{B}_1)}{\mu_0}
    +\dfrac{\vec{B}_0 \cdot \nabla\vec{B}_1}{\mu_0},
                    \label{eq_linMHD_momen} \\[0.1cm] 
    \dfrac{\partial \vec{B}_1}{\partial t}
&=& \nabla\times\left(\vec{v}_1\times \vec{B}_0 \right).
   					\label{eq_linMHD_Farad}                      
\end{eqnarray}
Evidently, the right-hand side (RHS) of Eq.~\eqref{eq_linMHD_momen}
    represents the linearized \Ampere\ force,
    the only restoring force ($\vec{f}$)
    in pressureless MHD.
By the second equal sign we show the nominal 
    decomposition of $\vec{f}$ into the magnetic pressure gradient force ($\vecfPG$)
   and magnetic tension force ($\vecfT$).
These further write
\begin{eqnarray}
&& \vecfPG = \fPGx\uvec{x} + \fPGy\uvec{y} 
         = -\dfrac{B_0}{\mu_0}
            \left(\dfrac{\partial B_{1z}}{\partial x} \uvec{x}
            	 +\dfrac{\partial B_{1z}}{\partial y} \uvec{y}
            \right), 
            \label{eq_vecfPG}\\
&& \vecfT = \fTx\uvec{x} + \fTy\uvec{y} 
         = \dfrac{B_0}{\mu_0}
            \left(\dfrac{\partial B_{1x}}{\partial z} \uvec{x}
            	 +\dfrac{\partial B_{1y}}{\partial z} \uvec{y}
            \right), 
            \label{eq_vecfT}
\end{eqnarray}
     where the non-contributing field-aligned components are excluded 
     (see the cylindrical study by \citealt{2009A&A...503..213G}).

We perform classic mode analysis for 3D motions hereafter, 
    kicking off with the Fourier ansatz,
\begin{equation}
	\label{eq_Fourier_ansatz}
	g_1 (x, y, z; t)
	= \Re\{\tilde{g}(x)\exp[-\imath (\omega t- k_y y -k_z z)]\},
\end{equation}
    where $g_1$ represents any linear perturbation,
    with $k_y$ ($k_z$) being real-valued out-of-plane (axial) wavenumbers.
We see the angular frequency $\omega$ as real-valued as well. 
In component form, Equations~\eqref{eq_linMHD_momen} and \eqref{eq_linMHD_Farad}
    then write 
\begin{eqnarray}
&&  \omega\tilde{v}_x 
= -\dfrac{B_0}{\mu_0 \rho_0}
     \left(k_z \tilde{B}_x + \imath \tilde{B}'_z\right), 
		\label{eq_Fourier_vx}	\\[0.2cm] 
&&  \omega\tilde{v}_y 
=  -\dfrac{B_0}{\mu_0\rho_0}
     \left(k_z \tilde{B}_y - k_y \tilde{B}_z\right), 
     	\label{eq_Fourier_vy}	\\[0.2cm] 
&&  \omega\tilde{B}_x 
=  -B_0 k_z \tilde{v}_x, 
	  	\label{eq_Fourier_Bx}	\\[0.2cm] 
&&  \omega\tilde{B}_y 
=  -B_0 k_z \tilde{v}_y, 
	    \label{eq_Fourier_By}	\\[0.2cm] 
&&  \omega\tilde{B}_z 
=  -\imath B_0 
	  \left(\tilde{v}_x'+ \imath k_y \tilde{v}_y\right), 
   \label{eq_Fourier_Bz} 
\end{eqnarray}
    where we adopt the shorthand notation $' \coloneqq \mathd/\mathd x$.
By ``mode'' we refer to a nontrivial solution, which is jointly characterized 
    by a mode frequency $\omega$ and  {an eigenvector} 
    $\{\tilde{v}_x, \tilde{v}_y, \tilde{B}_x, \tilde{B}_y, \tilde{B}_z\}$.
We see the wavenumbers $k_y$ and $k_z$ as independents, 
    taking equilibrium quantities to be parameters only.

\subsection{Fast waves in unbounded uniform media}
\label{sec_prob_sub_uniform}
This subsection examines an unbounded uniform medium
    for future reference. 
Now that $\rho_0 = \text{const}$, Fourier decomposition is allowed
    for the $x$-direction as well, enabling the quantity $\tilde{g}$ 
    in Eq.~\eqref{eq_Fourier_ansatz} to be expressible as
\begin{equation}
    \label{eq_FourierDecomp_uni}
    \tilde{g}(x) = \breve{g} \Exp{\imath k_x x}.
\end{equation}
Here $\breve{g}$ is a constant.
We focus on fast modes, or equivalently compressional \Alf\ waves, by 
    assuming $k_y\ne 0$.
Some explicit expressions for the restoring forces readily follow,
\begin{equation}
   \label{eq_forces_uni}
   \breve{\vec{f}}^{\rm PG}
=  \dfrac{B_0^2}{\mu_0} 
   \dfrac{\omega^2/\vA^2 - k_z^2}{\imath\omega} 
   \breve{\vec{v}}_{\perp}, 
\qquad 
   \breve{\vec{f}}^{\rm T}
=  \dfrac{B_0^2}{\mu_0} 
   \dfrac{k_z^2}{\imath\omega} 
   \breve{\vec{v}}_{\perp}
\coloneqq 
   \dfrac{\breve{\vec{f}}^{\rm PG}}{\Lambda}, 
\end{equation}
	where $\breve{\vec{v}}_{\perp} = \breve{v}_x \uvec{x} + \breve{v}_y \uvec{y}$.
It then follows that the parameter 
    \begin{equation}
    \label{eq_Lambda_uni}
    \Lambda = \dfrac{\omega^2}{k_z^2 \vA^2}-1    
    \end{equation}
    adequately quantifies how one force is related to the other. 

Some general properties for fast modes can be readily deduced.
To start, the DR reads
\begin{equation}
    \label{eq_DR_uni}
    \omega^2 = (k_x^2 + k_y^2 + k_z^2) \vA^2
             \coloneqq \kUniMedia^2 \vA^2, 
\end{equation}
    where a 3D wavevector 
    $\vec{\kUniMedia} = k_x \uvec{x}+ k_y \uvec{y} +k_z \uvec{z}$
    is introduced. 
Plugging Eq.~\eqref{eq_DR_uni} into Eq.~\eqref{eq_Lambda_uni} yields that
    \begin{equation}
        \label{eq_Lambda_uni2}
        \Lambda = \dfrac{k_x^2 + k_y^2}{k_z^2}.
    \end{equation}
This means that the magnetic pressure gradient force and 
    the tension force are always in-phase, thereby consistently complementing 
    each other to drive fast motions. 
Some subtlety arises for near-parallel propagation ($k_x^2 + k_y^2 \ll k_z^2$),
    in which case the tension force dominates such that fast modes become nearly 
    degenerate with shear \Alf\ waves. 
We proceed to define the phase and group velocities as
    \begin{equation}
    \label{eq_defVphVg_uni}
        \vec{v}_{\rm ph} \coloneqq \dfrac{\omega}{\kUniMedia} \uvec{\kUniMedia}, 
    \qquad
        \vec{v}_{\rm gr} \coloneqq \dfrac{\partial \omega}{\partial \vec{\kUniMedia}}
                        = \dfrac{\partial \omega}{\partial k_x} \uvec{x}
                         +\dfrac{\partial \omega}{\partial k_y} \uvec{y}
                         +\dfrac{\partial \omega}{\partial k_z} \uvec{z},
    \end{equation}
    where $\uvec{\kUniMedia} = \vec{\kUniMedia}/\kUniMedia$.
We assume $\{k_x, k_y, k_z, \omega\}$ to be non-negative without loss of generality.
The DR (Eq.~\eqref{eq_DR_uni}) then dictates that 
    \begin{equation}
        \vecvph = \vecvgr = \vA \uvec{\kUniMedia}, 
    \end{equation}
    namely fast modes are dispersionless. 
That fast modes are \Alf-like for near-parallel propagation is reflected in the fact that
   $\vec{v}_{\rm gr}$ is largely aligned with 
   the equilibrium field $\vec{B}_0 = B_0 \uvec{z}$.

\subsection{Kink modes in slab equilibria}
\label{sec_prob_sub_slab}
This subsection proceeds to examine a slab equilibrium,
    \begin{equation}
    \label{eq_rho_slab}
      \rho_0(x)
    = \left\{
       \begin{array}{ll}
          \rhoi,	    	& |x| < d, \\ [0.2cm]
          \rhoe,			& |x| > d,
       \end{array}
       \right.
    \end{equation} 
    where $d$ is the slab half-width. 
By ``internal'' (subscript ${\rm i}$) and ``external'' (subscript ${\rm e}$), 
    we will consistently refer to the equilibrium quantities inside and outside the slab, respectively.
In particular, the internal (external) \Alf\ speed, $\vAi$ ($\vAe$),
    is evaluated with the internal (external) density. 
Note that a step profile (Eq.~\eqref{eq_rho_slab})
    is adopted to avoid the resonant absorption of 3D kink motions in the \Alf\ continuum (see \citealt{2011SSRv..158..289G} for conceptual clarifications).
Note further that pressureless MHD does not allow slow motions per se. 
However, we avoid the single use of ``fast'' to describe
    the collective behavior of compressible motions
    (see \citealt{2009A&A...503..213G} for more;
    see also \citealt{2020A&A...641A.106G,2021A&A...646A..86G}).
Rather, such terms  {as ``$\vec{B}_0$-straddling'' or ``$\vec{B}_0$-same-side''} will be used
    to characterize 3D kink motions from the standpoint of group velocities.

We focus on trapped kink modes by supplementing Eqs.~\eqref{eq_Fourier_vx} to 
    \eqref{eq_Fourier_Bz} with appropriate boundary conditions (BCs). 
Only the half volume $x\ge 0$ needs to be considered for the resulting 
    boundary value problem (BVP).
The BC at the slab axis ($x=0$) is specified as
$ \tilde{v}_x'  = \tilde{v}_y 
= \tilde{B}_x'  = \tilde{B}_y = \tilde{B}_z = 0$, 
    while all Fourier amplitudes are required to vanish
    when $x \to \infty$.
All mode frequencies are real-valued.
Now that the BCs are homogeneous, some straightforward dimensional analysis
    yields that the mode frequencies can be formally written as
    \begin{eqnarray}
      \dfrac{\omega_j d}{\vAi} 
    = W_j(k_y d, k_z d | \rhoi/\rhoe)
    = W_j(k_y d, k_z d). 
    	\label{eq_omgFormal}
    \end{eqnarray}
By the second equal sign we recall that 
    the density contrast $\rhoi/\rhoe$ is seen as known.
The transverse order ($j=1, 2, \cdots$) numbers the mode frequencies
    at a given pair $[k_y, k_z]$ by increasing order.
We follow the convention that the transverse fundamental corresponds to $j=1$,
    while its overtones correspond to $j \ge 2$.
We detail only the cases $j=1$ and $j=2$, 
    noting that our analysis readily generalizes to any $j$. 
The subscript $j$ will be dropped for brevity hereafter, 
    unless confusions may arise.

Some generic properties ensue without solving the BVP.
One readily verifies that if $\omega$ is 
    a mode frequency for a given pair $[k_y, k_z]$, then so is $-\omega$.
Likewise, if $\omega$ is a mode frequency for $[k_y, k_z]$,
   then it remains so for the pairs $[-k_y, k_z]$, $[k_y, -k_z]$, and $[-k_y, -k_z]$. 
It therefore suffices to see $\omega$ as positive, and 
   consider only the quadrant $k_y>0, k_z>0$.  
Furthermore, Equations~\eqref{eq_Fourier_vx} to \eqref{eq_Fourier_Bz} 
    allow the Fourier amplitudes of the restoring forces 
   to be expressed as 
   (see Eqs.~\eqref{eq_vecfPG} and \eqref{eq_vecfT})
    \begin{equation}
    \label{eq_forces_slab}
       \tilde{\vec{f}}^{\rm PG}
    =  \dfrac{B_0^2}{\mu_0} 
       \dfrac{\omega^2/\vA^2 - k_z^2}{\imath\omega} 
       \tilde{\vec{v}}_{\perp}, 
    \qquad
       \tilde{\vec{f}}^{\rm T}
    =  \dfrac{B_0^2}{\mu_0} 
       \dfrac{k_z^2}{\imath\omega} 
       \tilde{\vec{v}}_{\perp}
    \coloneqq 
      \dfrac{\tilde{\vec{f}}^{\rm PG}}{\Lambda},  
    \end{equation}   
    where $\tilde{\vec{v}}_{\perp} = \tilde{v}_x \uvec{x} + \tilde{v}_y \uvec{x}$.
Equation~\eqref{eq_forces_slab} is formally identical to its counterpart in uniform media
    (Eq.~\eqref{eq_forces_uni}), the key difference being that the \Alf\ speed $\vA$ and the quantity $\Lambda$ need to be discriminated between the interior and exterior. 
We focus on the interior, supplementing $\Lambda$ with the subscript ${\rm i}$.
Evidently, 
    \begin{equation}
    \label{eq_LambdaI}
    \Lambda_{\rm i} = \dfrac{\omega^2}{k_z^2 \vAi^2}-1.    
    \end{equation}

Some specific results follow from explicit solutions to the BVP.
We proceed by defining
    \begin{eqnarray}
    &&  \kappa^2_{\rm i, e} 
    \coloneqq   
        k_z^2 - \dfrac{\omega^2}{v^2_{\rm Ai, e}},  
    	\label{eq_def_kappaie} \\
    &&  m^2_{\rm i, e} 
    \coloneqq  
        k_y^2 + \kappa^2_{\rm i, e} 
      = k_y^2 + k_z^2 - \dfrac{\omega^2}{v^2_{\rm Ai, e}}, 
    	\label{eq_def_mie}  \\
    && 	\myni^2 
    \coloneqq  -\mi^2 
    = \dfrac{\omega^2}{\vAi^2} - (k_y^2 + k_z^2). 
         \label{eq_def_ni}  
    \end{eqnarray}
We see both $\me^2$ and $\me$ as positive.
The external mode functions therefore write $\propto \Exp{-\me x}$,
    meaning that $\me$ can be taken as a measure of the capability
    for slabs to trap kink motions.
The signs of $\kappa^2_{\rm i, e}$ and $\mi^2$
    are unknown at this point.
We see kink modes as belonging to the ``surface'' (``body'') 
    subfamily when $\mi^2 >0$ ($-\myni^2 = \mi^2 <0$), 
    taking $\mi >0$ ($\myni >0$) without loss of generality.
Trapped 3D kink modes in our slab configuration 
    are known to obey the DR
    \citep[e.g.,][]{2007SoPh..246..213A,2021SoPh..296...95Y},
    \begin{equation}
    \label{eq_DR_coth}    
       \coth(\mi d)  
    =  \left(\dfrac{\me}{\mi}\right)
       \left(\dfrac{-\kappai^2}{\kappae^2}\right), 			
    \end{equation}
    or equivalently
    \begin{equation}
    \label{eq_DR_cot}
       \cot(\myni d) 
    = \left(\dfrac{\me}{\myni}\right)
      \left(\dfrac{-\kappai^2}{\kappae^2}\right).
    \end{equation}
We choose to work with Eq.~\eqref{eq_DR_coth} (Eq.~\eqref{eq_DR_cot})
    when handling surface (body) modes. 
Let $\vec{k} = k_y \uvec{y} + k_z \uvec{z}$.
This study pays special attention to both the phase and group velocities
    as defined by
    \begin{equation}
    \label{eq_defVphVg_slab}
        \vecvph \coloneqq \dfrac{\omega}{k} \uvec{k}, 
    \qquad
          \vecvgr 
       =  \vgy \uvec{y} + \vgz \uvec{z}
       \coloneqq 
          \dfrac{\partial \omega}{\partial k_y} \uvec{y}
         +\dfrac{\partial \omega}{\partial k_z} \uvec{z},
    \end{equation}
    where $\uvec{k} = \vec{k}/k$.

Of future use is the situation where $k_y = 0$.  
The DR of 2D kink modes is now textbook material, 
    writing \citep[see Sect.~5.5 in ][and references therein]{2019CUP_Roberts}
    \begin{equation}
    \label{eq_DR2D_cot}
       \cot(\myni d) 
    =  \dfrac{\myni}{\me}.
    \end{equation}
Mathematically, Equation~\eqref{eq_DR2D_cot} can be derived from its 3D counterpart
    by letting $k_y = 0$.
Equation~\eqref{eq_DR_cot} is more appropriate than Eq.~\eqref{eq_DR_coth}
    for this purpose, because 2D kink modes belong exclusively to 
    the body family. 
A series of cutoff axial wavenumbers $k_z^{(J)}$ are well known to arise,
    \begin{equation}
    \label{eq_defkzj}
      k_z^{(J)} d
    = \dfrac{J\pi}{\sqrt{\rhoi/\rhoe-1}}, 
    \end{equation}
    where $J=1, 2, \cdots$ in the 2D context. 
Transverse fundamentals ($j=1$) exist for any $k_z >0$. 
Taking the first overtone ($j=2$) as example, 
    by ``cutoff'' we then mean that trapped modes are allowed
    only when $k_z > k_z^{(j-1)} = k_z^{(1)}$
    \citep[e.g.,][]{1995SoPh..159..213N,2013ApJ...767..169L}.
The solutions to Eq.~\eqref{eq_DR2D_cot}
    will be denoted by $\omgtwoD = \omgtwoD (k_z)$ for clarity.
Equation~\eqref{eq_defkzj} is also involved in our examination
    of 3D kink modes, which nonetheless require that half integers ($J=1/2, 3/2, \cdots$)     be addressed.

\section{Results}
\label{sec_results}
This study is intended to demonstrate how classic mode analysis
    can be better placed in the context of impulsively excited waves. 
Regarding the DR (Eq.~\eqref{eq_DR_coth} or equivalently Eq.~\eqref{eq_DR_cot}),
    it should be ideal that one deduces those generic properties 
    that are insensitive to the density contrast $\rhoi/\rhoe$.
However, the transcendental nature of the DR does not allow 
    a completely analytical treatment.   
We will adopt a fixed $\rhoi/\rhoe =3$, 
    a value relevant for, say, active region loops \citep[e.g.,][]{2004ApJ...600..458A},
    polar plumes \citep[e.g.,][]{2011A&ARv..19...35W},
    and streamer stalks \citep[e.g.,][]{2011ApJ...728..147C}.
We will solve the DR with standard root-finders for this specific $\rhoi/\rhoe =3$,
    and build a more generic picture by proceeding analytically in various limiting cases.  

\subsection{Generic characterization}
\label{sec_results_generic}
This subsection characterizes kink modes by 
    inspecting the form of the DR (Eqs.~\eqref{eq_DR_coth} and \eqref{eq_DR_cot}).
The mode frequencies $\omega$ for a given transverse order $j$ 
    then map to a dispersion sheet in the 3D space
    $[k_y, k_z, \omega]$.
We choose to present a dispersion sheet 
    by showing some of its cuts through constant values of $k_z$.
A dispersion curve then results, expressing $\omega$ as a function of $k_y$
    for a given $k_z$.
One readily recognizes from Eq.~\eqref{eq_DR_cot} that the surfaces 
    $\kappai^2=0$, $\kappae^2=0$, $\me=0$, and $\myni = J\pi~(J=0, 1/2, 1, 3/2, \cdots)$
    are important in organizing the dispersion sheets; 
    these surfaces are where the LHS or RHS of Eq.~\eqref{eq_DR_cot} changes sign. 
It follows that the intersections of these surfaces are likely to be important as well.
In particular, the intersection between the surfaces $\myni=J\pi$ and $\me=0$ 
    satisfies 
    \begin{equation}
        (k_y^2 + k_z^2) d^2       = \dfrac{(J\pi)^2}{\rhoi/\rhoe -1},
        \qquad
        (J=1/2, 1, 3/2, \cdots)
    \end{equation}
    which defines a set of circles in the $k_y-k_z$ plane. 
The reason for $k_z^{(J)}$ (see Eq.~\eqref{eq_defkzj}) to be relevant
    is then that $k_z^{(J)}$ locates where these circles intersect $k_y=0$. 
The quantity $k_z^{(J)}$ evaluates to 
    \begin{equation}
    \label{eq_kzj_rhoie3}
        k_z^{(1/2)}d = 1.11, \quad
        k_z^{(1)}d   = 2.22, \quad
        k_z^{(3/2)}d = 3.33, \cdots
    \end{equation}
    for the chosen $\rhoi/\rhoe=3$.

\begin{figure*}
\centering
\includegraphics[width=.99\textwidth]{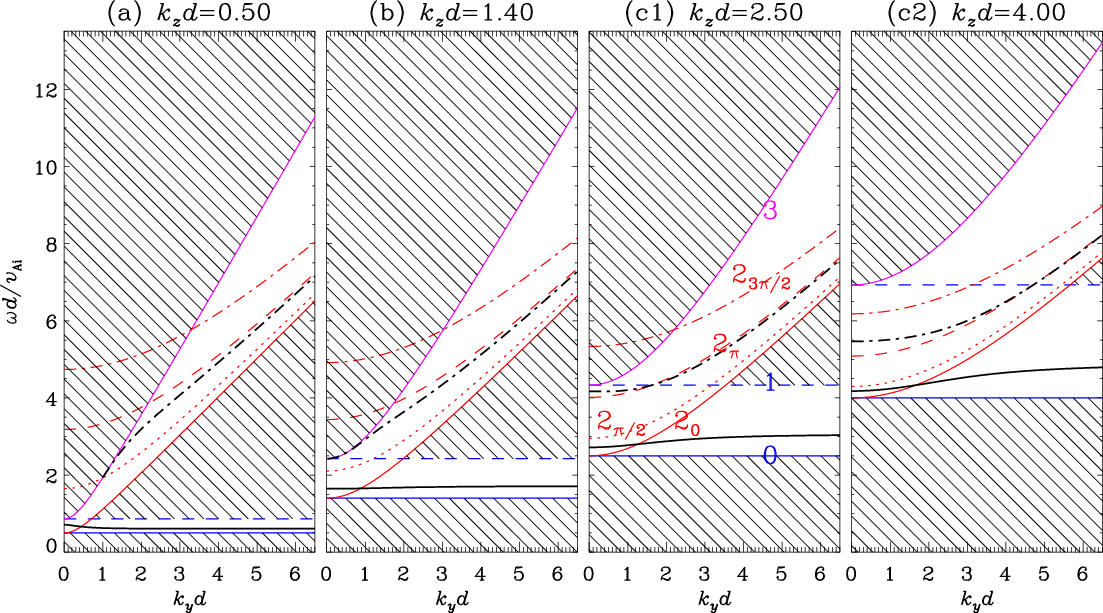}
\caption{
Characterization of trapped 3D kink modes in solar coronal slabs.
A density contrast of $\rhoi/\rhoe=3$ is chosen for illustrative purposes.
Plotted in each panel are the dependencies on the 
    out-of-plane wavenumber ($k_y$)
	of the mode frequency of the transverse fundamental 
	(with transverse order $j=1$, the black solid  curve) 
	and the first overtone ($j=2$, black dash-dotted).
The hatched portions in any $k_y-\omega$ plane represent 
    where trapped modes are forbidden, with four curves serving as
    the relevant borders (labeled in panel~c1). 
Curves~$0$, $1$, $2_0$, and $3$ correspond to
     where $\kappai^2 \coloneqq k_z^2 - \omega^2/\vAi^2=0$,
     $\kappae^2 \coloneqq k_z^2 - \omega^2/\vAe^2=0$,
     $\myni^2 \coloneqq \omega^2/\vAi^2 - (k_y^2+k_z^2) =0$,
     and $\me^2 \coloneqq k_y^2+k_z^2 - \omega^2/\vAe^2=0$, 
     respectively.
The curves labeled $2_{J\pi}$ ($J=1/2, 1, 3/2$) further correspond to where
     $\myni = J\pi$.
These curves pertain to the characterization of the $j=1$ and $j=2$ modes, 
     whose dispersive behavior is typified by different panels where the
     dimensionless axial wavenumber ($k_z d$) increases sequentially.
See text for more details.      
 }
\label{fig_charSols} 
\end{figure*}

Figure~\ref{fig_charSols} shows a series of $k_y - \omega$ planes 
    for an increasing sequence of
	dimensionless axial wavenumbers ($k_z d$) as indicated.
The associated mode frequency $\omega$ of the transverse
    fundamental (first overtone) 
    is shown by the black solid (dash-dotted) curve 
    for each $k_z$.
A number of curves are further displayed for reference, 
    with the color and linestyle being consistent across all panels. 
These are also labeled (see Fig.~\ref{fig_charSols}c1) to ease our description.    
Specifically,
\begin{itemize}
    \item Line $0$, the blue solid  line, corresponds to $\kappai^2 = 0$. 
    \item Line $1$, the blue dashed line, corresponds to $\kappae^2 = 0$.
    \item Curve $3$, the upper red solid curve, corresponds to $\me = 0$.
    \item Curve $2_0$, the lower red solid curve, corresponds to $\myni=0$.
    \item Curves $2_{J\pi}$ with $J=[1/2, 1, 3/2]$, correspond to $\myni=J\pi$ and are shown
        by the red dottted, dashed, and dash-dotted curves, respectively.
\end{itemize}
Above all, these curves are important for showing that trapped kink modes
    are allowed only in the non-hatched portion 
    in a $k_y - \omega$ plane.
The portion bounded by curve $3$ from below does not allow trapped modes 
    by definition, given that $\me^2 <0$ therein.
That trapped modes are prohibited in the other two hatched portions is because 
    there is always a mismatch between the signs of the
    LHS and RHS of Eq.~\eqref{eq_DR_coth}.

The specific value of $k_z$ does not qualitatively impact
    the transverse fundamental in the following aspects.
Firstly, the transverse fundamental is present 
    for any $k_z>0$ and $k_y \ge 0$.
Secondly, the fundamental always transitions from a body 
    to a surface type as $k_y$ increases, the dividing line being 
    $\myni=0$ (or equivalently $\mi^2 = 0$).
Thirdly, the dispersion curve of the fundamental  always lies in 
    the horizontal stripe bounded by lines $0$ and $1$, and is further bounded
    from above by curve $2_{\pi/2}$ when the fundamental is of the body type.
The reason is simply that the LHS of Eq.~\eqref{eq_DR_coth} or Eq.~\eqref{eq_DR_cot}
    needs to be positive definite.
    
The behavior of the first overtone may be qualitatively different for 
    different values of $k_z$.
This is so despite that the first overtone is always of the body type.    
Three regimes need to be discriminated regarding how the $2_{J\pi}$ (with $J=1/2, 1, 3/2$)
    curves are positioned with respect to the stripe between lines $0$ and $1$.
\begin{itemize}
    \item Regime I with $0 < k_z < k_z^{(1/2)}$ as typified
        by Fig.~\ref{fig_charSols}a. 
    All three $2_{J\pi}$ curves lie above the horizontal stripe. 
    One readily verifies that the dispersion curve starts from the intersection 
        between curves $3$ and $2_{\pi/2}$, running always between curves
        $2_{\pi}$ and $2_{\pi/2}$. 
    Of relevance is then some cutoff value of $k_y$, only beyond which
        is the first overtone allowed.
    \item Regime II with $k_z^{(1/2)} < k_z < k_z^{(1)}$ as typified
        by Fig.~\ref{fig_charSols}b. 
    Only the $2_{\pi/2}$ curve extends into the horizontal stripe. 
    One readily verifies that the dispersion curve starts from the intersection 
        between curve $3$ and line $1$.
    The pair $[k_y,  \omega]$ reads $[0, k_z \vAe]$ at this intersection, 
        where the transverse overtone is not allowed per se because $\me = 0$.
    The dispersion curve therefore lies entirely outside the horizontal stripe,
        running between curves $2_{\pi}$ and $2_{\pi/2}$.
    \item Regime III with $k_z > k_z^{(1)}$.
    Figures~\ref{fig_charSols}c1 and \ref{fig_charSols}c2 further typify what happens
        when $k_z$ is below and above $k_z^{(3/2)}$, respectively. 
    Curve $2_{\pi}$ extends into the horizontal stripe in the former, while
        curve $2_{3\pi/2}$ also does so in the latter.
    However, the dispersion curve is qualitatively the same in the following sense.     
    One readily verifies that the intersection between curve $2_{\pi}$ and line $1$
        always lies on the dispersion curve, which therefore can be divided into two portions.
    The portion in the horizontal stripe is always bounded by curve $2_{\pi}$ from below,
        and never touches curve $2_{3\pi/2}$ or line $1$. 
    On the other hand, the portion outside the horizontal stripe always runs 
        between curves $2_{\pi}$ and $2_{\pi/2}$.
\end{itemize}
All these properties follow from the inspection of the signs of the LHS and RHS 
    of Eq.~\eqref{eq_DR_cot}.
Similar conclusions can therefore be inferred for overtones of higher transverse order ($j$).
All overtones belong exclusively to the body type, even though some quantitative difference
    from the first overtone does arise for $j\ge 3$ regarding how the associated dispersion curve is positioned in the $k_y-\omega$ plane.
Take the second overtone $j=3$ for instance, by difference we mean that
    curves $2_{J\pi}$ with $J=3/2, 2, 5/2$ will be relevant simply because
    the $\cot$ function is $\pi$-periodic.

\subsection{Transverse fundamental}
\label{sec_results_fund}
\begin{figure}
\centering
\includegraphics[width=.99\columnwidth]{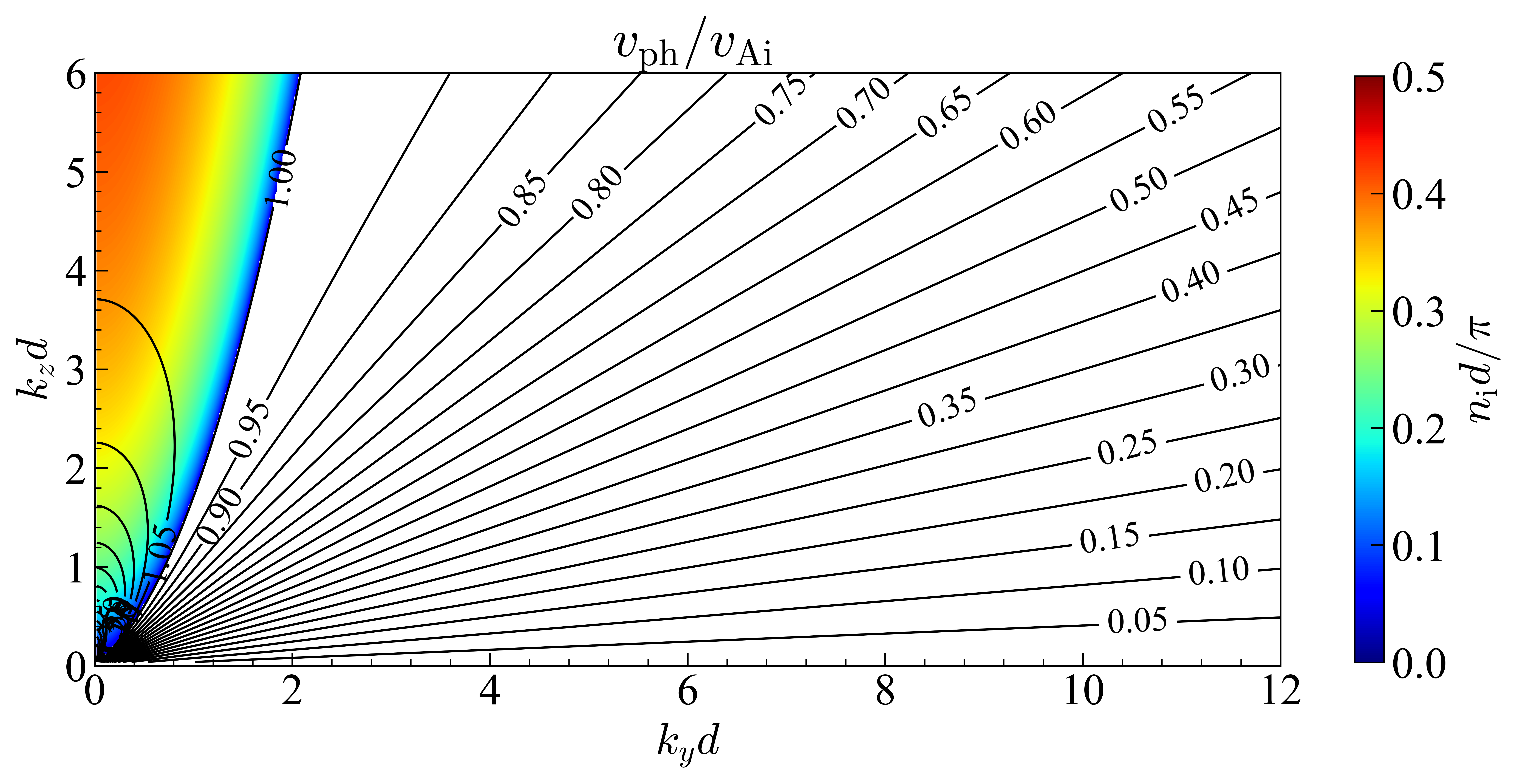}
\caption{
Transverse fundamental kink mode 
    in a coronal slab with a density contrast $\rhoi/\rhoe=3$.
The dependence on $[k_y, k_z]$  
    of the phase speed $\vph=\omega/\sqrt{k_y^2+k_z^2}$
    is shown by the equally spaced contours.
Trapped modes are allowed for all $[k_y \ge 0, k_z >0]$, 
    with both body ($\myni^2 > 0$) and surface ($\myni^2 < 0$)
    modes being relevant. 
The color map represents the parameter $\myni$,
    and is restricted to body modes. 
 }
\label{fig_fund_vph} 
\end{figure}

\begin{figure}
\centering
\includegraphics[width=.99\columnwidth]{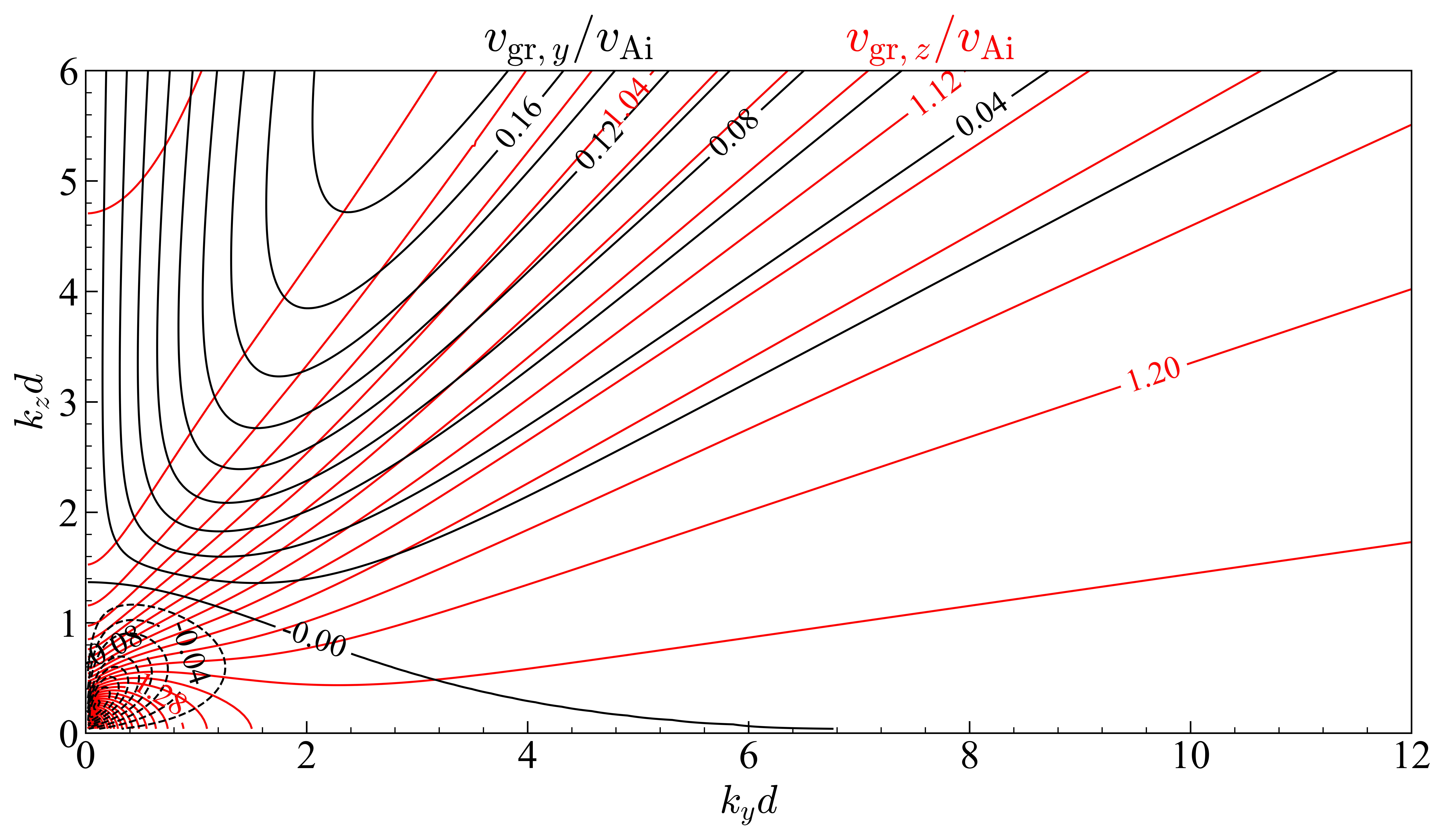}
\caption{
Transverse fundamental kink mode 
    in a coronal slab with a density contrast $\rhoi/\rhoe=3$.
Shown are the dependencies on $[k_y, k_z]$  
    of the $y$-component (the black contours)
    and the $z$-component (red) of the group velocity. 
All contours are equally spaced.     
}
\label{fig_fund_vgr} 
\end{figure}

This subsection is devoted to the transverse fundamental.
Figure~\ref{fig_fund_vph} displays the distribution in the 
    $k_y-k_z$ plane of the phase speed as equally spaced contours.
Also plotted are the filled contours of the parameter $\myni$,
    to evaluate which we now see $\omega$ in Eq.~\eqref{eq_def_ni}
    as the mode frequency.
Note that $\myni$ makes sense only for body modes ($\myni^2 >0$), 
    meaning that the portion not occupied by the color map
    corresponds to surface modes. 
Figure~\ref{fig_fund_vgr} further plots how the 
         $y$-component ($\vgy$, black contours)
    and  $z$-component ($\vgz$, red)
    of the group velocity depend on $[k_y, k_z]$.

Some analytical progress can be made to help digest 
    Figs.~\ref{fig_fund_vph} and \ref{fig_fund_vgr}. 
Consider near-parallel propagation first
    ($k_y^2 \ll k_z^2$), and start with the particular case $k_y=0$.
It is well documented for these 2D modes \citep[e.g.,][Chapter~5]{2019CUP_Roberts}
    that the phase speed $\omgtwoD(k_z)/k_z$
    decreases monotonically with $k_z$, 
    approaching $\vAe$ ($\vAi$) when $k_z d \to 0$ ($k_z d \to \infty$).
This behavior is what one sees along $k_y = 0$ in Fig.~\ref{fig_fund_vph}.    
One further deduces that
	\begin{eqnarray}
	&&          \omgtwoD(k_z) 
          \approx k_z \vAe \sqrt{1-(\rhoi/\rhoe-1)^2 (k_z d)^2}, 
	\quad k_z d \ll 1,
		    \label{eq_fund_ky0_kzsmall} \\
	&& \omgtwoD(k_z) 
		\approx k_z \vAi 
		\sqrt{1+\left(\pi/2\right)^2 (k_z d)^{-2}},	
		\quad k_z d \gg 1,
		\label{eq_fund_ky0_kzlarge} 
	\end{eqnarray}
	 by specializing Eqs.~(18) and~(16) in \citet{2013ApJ...767..169L}
	    to pressureless MHD.
Note that $\myni$ evaluates to $\sqrt{(\omgtwoD/\vAi)^2 - k_z^2}$ for 2D modes,
    and is itself a monotonically increasing function of $k_z$
    (see Fig.~\ref{fig_fund_vph}).
Equations~\eqref{eq_fund_ky0_kzsmall} and \eqref{eq_fund_ky0_kzlarge} 
    then help show that $\myni d \to 0$ for $k_z d \to 0$
    and $\myni d \to \pi/2$ when $k_z d \to \infty$. 

Now consider the situation $0 < k_y^2/k_z^2 \ll 1$.
It follows from \citet[][Eq.~(9)]{2023MNRAS.518L..57L} that
	\begin{equation}
	\label{eq_fund_smallKY}
	        \omega(k_y, k_z)
	\approx \omgtwoD 
			\left[1+\dfrac{1}{2}
			        \dfrac{\metwoD d -1}
			             {(k_z d)^2 + (\metwoD d)(\omgtwoD d/\vAi)^2}  
			        (k_y d)^2
			\right],      
	\end{equation}
    where $\omgtwoD=\omgtwoD(k_z)$
    and $(\metwoD)^2 \coloneqq k_z^2 - (\omgtwoD)^2/\vAe^2$.
Equation~\eqref{eq_fund_smallKY} explicitly shows that the correction to 
    $\omgtwoD$ is only quadratic in $k_y$, meaning that 
    $\vgy = \partial\omega/\partial k_y \to 0$
    when $k_y d\to 0$ (see Fig.\ref{fig_fund_vgr}).
More importantly, one notices from 
    Eqs.~\eqref{eq_fund_ky0_kzsmall} and \eqref{eq_fund_ky0_kzlarge}
    that $\metwoD \propto k_z^2$ for $k_z d \ll 1$ and
         $\metwoD \propto k_z$ for $k_z d \gg 1$.
Overall, $\metwoD$ turns out to be a monotonically increasing function of $k_z$. 
One then recognizes the existence of some critical axial wavenumber $k_z^{\rm 2D, c}$,
    which renders $\metwoD-1$ negative (positive) 
    when $k_z < k_z^{\rm 2D, c}$ ($k_z > k_z^{\rm 2D, c}$).  
Let $\mathcal{C}$ denote the coefficient in front of $(k_y d)^2$ 
    in Eq.~\eqref{eq_fund_smallKY}.
It then follows that $\mathcal{C}$ reverses its sign
    when $k_z d$ passes through $k_z^{\rm 2D, c} d$.
Somehow Fig.~\ref{fig_fund_vph} indicates that both $\vph$ and $\myni$
    tend to decrease with $k_y$ at any given $k_z$, demonstrating no signature of
    $k_z^{\rm 2D, c} d$.
This behavior can be explained by Eqs.~\eqref{eq_defVphVg_slab} and \eqref{eq_def_ni}
    where $\vecvph$ and $\myni$ are defined; 
    the direct $k_y$-dependencies tend to dominate 
    the indirect $k_y$-dependencies through $\omega$.
However, $k_z^{\rm 2D, c} d$ plays an important role for $\vgy$ and $\vgz$.
One sees that $\vgy$ tends to decrease to negative values with 
    increasing $k_y$ for small $k_z$, whereas
    $\vgy$ increases to positive values when $k_y$ increases from zero
    for large $k_z$.
This is a direct consequence of Eq.~\eqref{eq_fund_smallKY}, which actually helps one
    to deduce $k_z^{\rm 2D, c} d \approx 1.38$ from Fig.~\ref{fig_fund_vgr}
    by locating where $\vgy$ changes sign for $k_y \to 0$.
The quantity $k_z^{\rm 2D, c}$ is relevant for $\vgz$ as well, to show which
    we focus on the range $k_z < k_z^{\rm 2D, c}$.
One sees that $\vgz$ tends to decrease (increase) with $k_y$ 
    when $k_z$ is fixed at some small (large) value.
This is readily understandable if one notices the relevance of
    $\mathd\mathcal{C}/\mathd k_z$ when evaluating 
    $\partial \omega/\partial k_z$
    with Eq.~\eqref{eq_fund_smallKY}.

We now address the situation $k_y^2/k_z^2 \gg 1$ where
     sufficiently oblique modes can be safely taken as surface modes.
It is known that \citep[][]{1998JPSJ...67.2322T, 2021SoPh..296...95Y}
      \begin{equation}
      \label{eq_omg_fund_largeKY}
                \omega(k_y, k_z) 
        \approx k_z C(k_y), 
        ~~\text{with}~~
           \dfrac{C^2(k_y)}{\vAi^2}
		=  \dfrac{1+\tanh(k_y d)}{\rhoe/\rhoi+\tanh(k_y d)}
       \end{equation}   
    approximately solves Eq.~\eqref{eq_DR_coth}.
The more restrictive situation $k_y d \to \infty$ is well studied,
    yielding $C \to C_{\rm kink} = \vAi\sqrt{2/(\rhoe/\rhoi+1)}$
    \citep[e.g.,][]{1978ApJ...226..650I,1992SoPh..138..233G,1995JPlPh..54..129R}.
One sees from Fig.~\ref{fig_fund_vph} that the contours of the phase speed $\vph$ 
    largely become
    radially directed in the $k_y-k_z$ plane.
This behavior can be explained by the simplest approximation
    $\omega \approx k_z C_{\rm kink}$, which yields that 
    $\vph \approx C_{\rm kink}/\sqrt{1+(k_y/k_z)^2}$.
Furthermore, this simplest approximation also helps understand the overall
    tendencies for $\vgy$ to be small and for $\vgz$ to approach $C_{\rm kink}$
    when $k^2_y/k^2_z \gg 1$ (see Fig.~\ref{fig_fund_vgr}). 
However, some subtlety remains regarding the $k_y$-dependence of
    $\vgy$ or $\vgz$ even if the improved approximation 
    (Eq.~\eqref{eq_omg_fund_largeKY}) is employed.
For instance, one expects with Eq.~\eqref{eq_omg_fund_largeKY} that
    $\vgy$ approaches zero from below given that $C(k_y)$ decreases
    monotonically with $k_y$.
Figure~\ref{fig_fund_vgr}, however, indicates that $\vgy >0$ for sufficiently 
    large $k_y$.
Likewise, Eq.~\eqref{eq_omg_fund_largeKY} suggests that $\vgz$ should approach
    $C_{\rm kink}$ from above.
This contradicts the behavior of $\vgz$ at large $k_y$.
It turns out that Eq.~\eqref{eq_omg_fund_largeKY} needs to be improved to address
    the $k_y$-dependence for those ranges of $k_y$ where $\tanh(k_y d)$ varies only slowly.
Writing 
    \begin{equation}
    \label{eq_omg_fund_largeKY_mine}
        \omega \approx k_z C(k_y) \left(1-Q \dfrac{k_z^2}{k_y^2}\right),    
    \end{equation}
     we proceed to Taylor-expand all terms on both sides of the DR (Eq.~\eqref{eq_DR_coth}) 
     to the lowest-order in $k_z^2/k_y^2$.
The coefficient $Q$ then results from some algebra, reading
      \begin{equation}
      \label{eq_omg_fund_largeKY_Q}
          Q
        = \dfrac{1}{4}
          \dfrac{(1-\rhoe/\rhoi)^2}
                {[\rhoe/\rhoi+\tanh(k_y d)]^2}
          \left\{
             \tanh(k_y d) - (k_y d)[1-\tanh(k_y d)]   
          \right\}.      
      \end{equation}
For the chosen $\rhoi/\rhoe=3$, Equation~\eqref{eq_omg_fund_largeKY_mine} proves
    to be remarkably accurate when $k_y/k_z \gtrsim 2$.
Consequently, one readily explains the large-$k_y$ behavior
    of $\vgy$ or $\vgz$ by noting that 
    \begin{subequations}
        \begin{align}
        &         Q
          \approx \dfrac{1}{4}
                  \dfrac{(1-\rhoe/\rhoi)^2}
                        {(1+\rhoe/\rhoi)^2},     \quad 
                  \omega 
          \approx k_z C_{\rm kink}
                       \left(1-Q \dfrac{k_z^2}{k_y^2}\right),    \\
        &         \dfrac{\vgy}{C_{\rm kink}}
          \approx 2 Q \dfrac{k_z^3}{k_y^3},                \quad 
                  \dfrac{\vgz}{C_{\rm kink}}
          \approx 1-3 Q \dfrac{k_z^2}{k_y^2}
        \end{align}
    \end{subequations}
    for sufficiently large $k_y d$.

\begin{figure}
\centering
\includegraphics[width=.99\columnwidth]{f04_updated.png}
\caption{
Transverse fundamental kink mode 
    in a coronal slab with a density contrast $\rhoi/\rhoe=3$.
Shown are the dependencies on $[k_y, k_z]$  
    of (a) the angle between the group velocity $\vecvgr$
           and the $z$-direction,
    and (b) the angle between the group velocity $\vecvgr$
           and the phase velocity $\vecvph$.
The angles are in degrees and measured counterclockwise from $\vecvgr$ to 
    $\hat{\vec{e}}_z$ or $\vecvph$.
Superimposed are the filled contours of $\Lambdai$,
    the ratio of the magnetic pressure gradient force
    to the magnetic tension force.
All contours are equally spaced.
The angle information allows the fundamental to be 
    classified into two regimes as labeled 
    in panel~c.
See text for details.     
 }
\label{fig_fund_angles} 
\end{figure}

Figure~\ref{fig_fund_angles} gathers the directional information
    of the group ($\vecvgr$) and phase velocities ($\vecvph$),
    aiming to categorize transverse fundamental kink modes
    on the $k_y-k_z$ plane.
Plotted are the $[k_y, k_z]$-dependencies of 
    (a) the angle ($\angle(\vecvgr, \uvec{z}$)) between $\vecvgr$
           and the $z$-direction,
    and (b) the angle ($\angle(\vecvgr, \vecvph$) between the two vector
        velocities themselves.
The angles are measured counterclockwise from $\vecvgr$, 
    the readings being in degrees.
All contours are equally spaced.
Also shown by the filled contours is the parameter $\Lambdai$,
    which is recalled to be the ratio of the magnetic pressure gradient force
    to the magnetic tension force (see Eq.~\eqref{eq_LambdaI}).

Consider the quantity $\Lambdai$ first.     
By definition, $\Lambdai$ is positive definite for the transverse fundamental
    and overtones alike, given that all trapped modes ensure 
    $\kappai^2 = k_z^2 - \omega^2/\vAi^2 <0$
    (see Fig.~\ref{fig_charSols}).
Figure~\ref{fig_fund_angles}a further indicates that $\Lambdai$
    tends to be smaller than unity for the majority of the $[k_y, k_z]$ combinations,
    with the portion $\Lambdai > 1$ restricted to the lower left corner.
This can be understood with the approximate expressions that were presented
    previously.
The expressions for 2D modes ($k_y = 0$, 
    Eqs.~\eqref{eq_fund_ky0_kzsmall}
    and~\eqref{eq_fund_ky0_kzlarge}) 
    suggest that $\Lambdai \to \rhoi/\rhoe-1$ when $k_z d \ll 1$
    and $\Lambdai \to 0$ when when $k_z d \to \infty$.
Addressing a small $k_y^2/k_z^2$,
    Equation~\eqref{eq_fund_smallKY} further suggests that
    $\Lambdai$ decreases (increases) with $k_y$
    for relatively small (large) values of $k_z$ 
    that ensure a negative (positive) $k_y^2$-correction. 
Likewise, one deduces from Eq.~\eqref{eq_omg_fund_largeKY_mine}
    that 
    \begin{equation}
    \label{eq_fund_Lambdai_largeky}
    \Lambdai \to \dfrac{C_{\rm kink}^2}{\vAi^2}-1 
               =\dfrac{1-\rhoe/\rhoi}{1+\rhoe/\rhoi} 
               < 1,
    \end{equation}
    when $k_y^2/k_z^2 \to \infty$.
Actually, one sees from Fig.~\ref{fig_fund_angles}a that
    $\Lambdai$ is nearly uniform for sufficiently large $k_y$; 
    the subtle $k_y-$ (and $k_z$-) dependencies of $C(k_y)$
    and the $Q$ term play only a marginal role in determining the gross
    distribution of $\Lambdai$.

We now attempt to place the angle distributions in
    the context of $\Lambdai$. 
For immediate future reference, we classify trapped 3D kink modes
    into the following three regimes, noting that $\vgz$ is positive definite
     {in our context. 
    \begin{itemize}
        \item ``$\vec{B}_0$-straddling'', by which we mean the situation where
                $\vgy<0$ (i.e., $\vecvgr$ and $\vecvph$ straddle $\vec{B}_0$).
        \item ``$\vec{B}_0$-same-side A'', by which we mean the situation where $\vgy>0$ and 
                $|\angle(\vecvgr, \uvec{z})|<15^\circ$.   
        \item ``$\vec{B}_0$-same-side F'', by which we refer to the situation
                where $\vgy>0$, $|\angle(\vecvgr, \uvec{z}|>15^\circ$, and
                      $|\angle(\vecvgr, \vecvph)|<15^\circ$.
    \end{itemize}
Note that by ``$\vec{B}_0$-same-side'' we mean $\vecvgr$ and $\vecvph$ 
    sit on the same side of $\vec{B}_0$.}    
This classification scheme largely comes from the analogy
    with 3D waves ($k_y > 0$) in unbounded uniform media
    (see Sect.~\ref{sec_prob_sub_uniform}), with the threshold
    angle arbitrarily chosen to be some small value ($15^\circ$ here).
Note that the waves examined in Sect.~\ref{sec_prob_sub_uniform},
    broadly called ``fast'' therein, need to be categorized into 
     {the ``$\vec{B}_0$-same-side A'' and ``$\vec{B}_0$-same-side F'' subgroups here.}
Equation~\eqref{eq_Lambda_uni2} then offers an unambiguous association
    of this categorization with the force ratio; 
    the  {``$\vec{B}_0$-same-side A''} subgroup arises when the force ratio 
    $\Lambda < \tan^2(15^\circ)\approx 0.07$,
    whereas the  {``$\vec{B}_0$-same-side F''} subgroup arises when the opposite is true.
 {The letters ``A'' and ``F'' are meant to distinguish the
    \Alf-like modes and the genuinely fast-like modes.} 
The  {``$\vec{B}_0$-straddling'' subgroup is absent in Sect.~\ref{sec_prob_sub_uniform},
    but bears substantial resemblance to slow waves
    in unbounded uniform media with non-vanishing puressue} \citep[e.g.,][Sect.~2.7]{2019CUP_Roberts}.
Figures~\ref{fig_fund_angles}a and \ref{fig_fund_angles}b
    indicate that transverse fundamental kink modes 
     {qualify as ``$\vec{B}_0$-same-side A''}
    in the majority of the $k_y-k_z$ plane.
This is true even for highly oblique propagation ($k_y^2/k_z^2 \gg 1$), 
    despite that the force ratio $\Lambdai$ is only marginally small
    ($1/2$ for the chosen $\rhoi/\rhoe=3$, see Eq.~\eqref{eq_fund_Lambdai_largeky}).
In this regard, our classification scheme agrees with the force-ratio-based
    argument by \citet{2009A&A...503..213G} in that transverse fundamental kink modes are largely \Alf-like, despite that cylindrical geometry was examined therein. 
More striking is the  {``$\vec{B}_0$-straddling'' subgroup}, 
    which shows up at the lower left
    corner in the $k_y-k_z$ plane.
By striking we mean that pressureless MHD does not allow slow waves
    in textbook sense as far as unbounded uniform media are concerned.
However, these motions do make it into Fig.~\ref{fig_fund_angles}, thereby
    highlighting the intricacies that transverse structuring brings
    into wave studies. 
For clarity, we collect our classification in Fig.~\ref{fig_fund_angles}c
    and label the subgroups accordingly.
One sees that the force ratio $\Lambdai \gtrsim 1$ 
    in the  {``$\vec{B}_0$-straddling'' corner}.

\subsection{First transverse overtone}
\label{sec_results_overtone}
\begin{figure}
\centering
\includegraphics[width=.99\columnwidth]{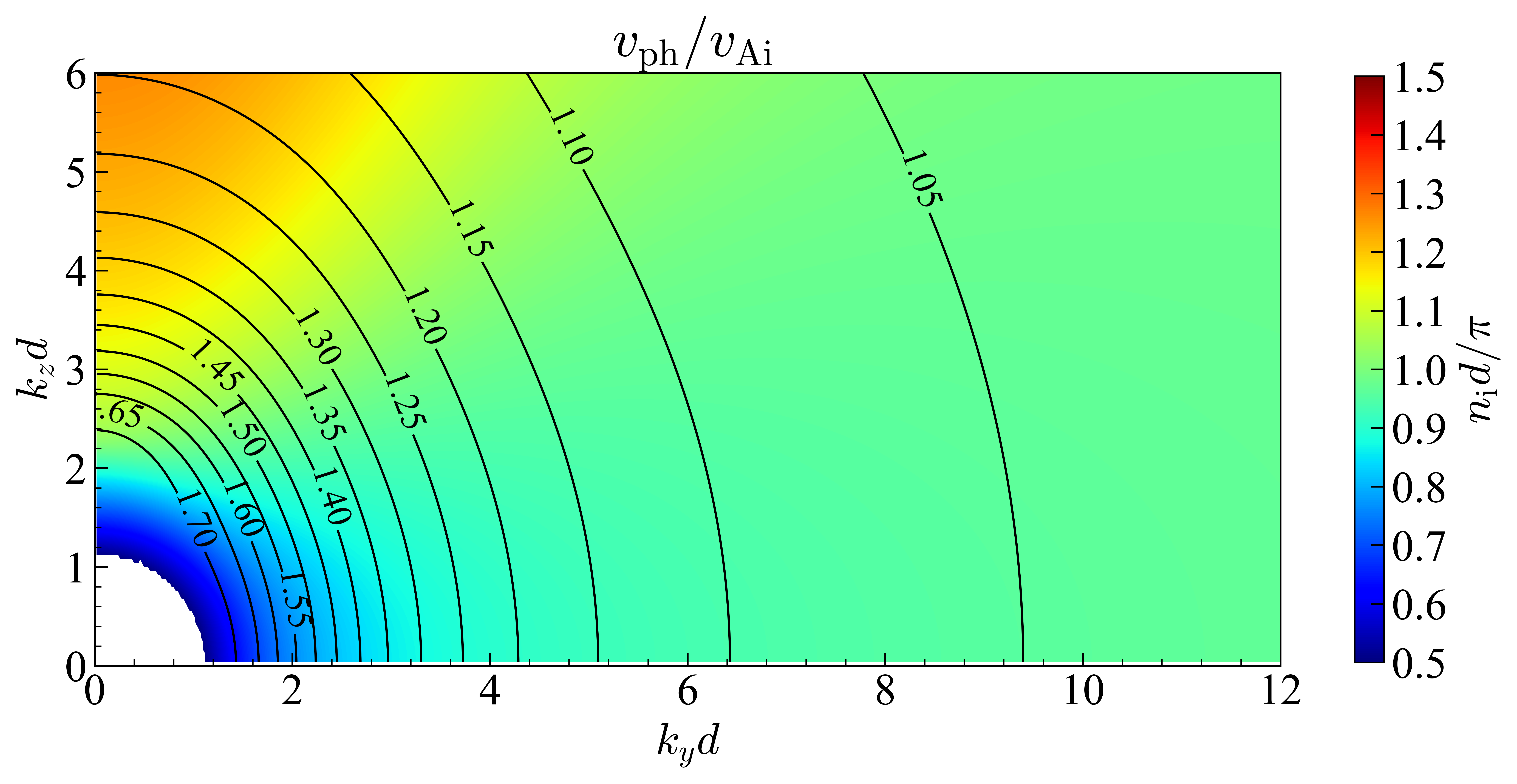}
\caption{
Similar to Fig.~\ref{fig_fund_vph} but for 
	the first transverse overtone.
Trapped 3D modes are forbidden within the circle
    $d\sqrt{k_y^2 + k_z^2} = k_z^{(1/2)} d = (\pi/2)/\sqrt{\rhoi/\rhoe -1}$,
    which reads $1.11$ for the chosen $\rhoi/\rhoe=3$.
Trapped 2D modes are further prohibited for $k_z^{(1/2)} < k < k_z^{(1)} = 2k_z^{(1/2)} $ 
    along $k_y=0$.
Trapped modes are exclusively of the body type ($\myni^2 >0$). 
 }
\label{fig_overtone_vph} 
\end{figure}

\begin{figure}
\centering
\includegraphics[width=.99\columnwidth]{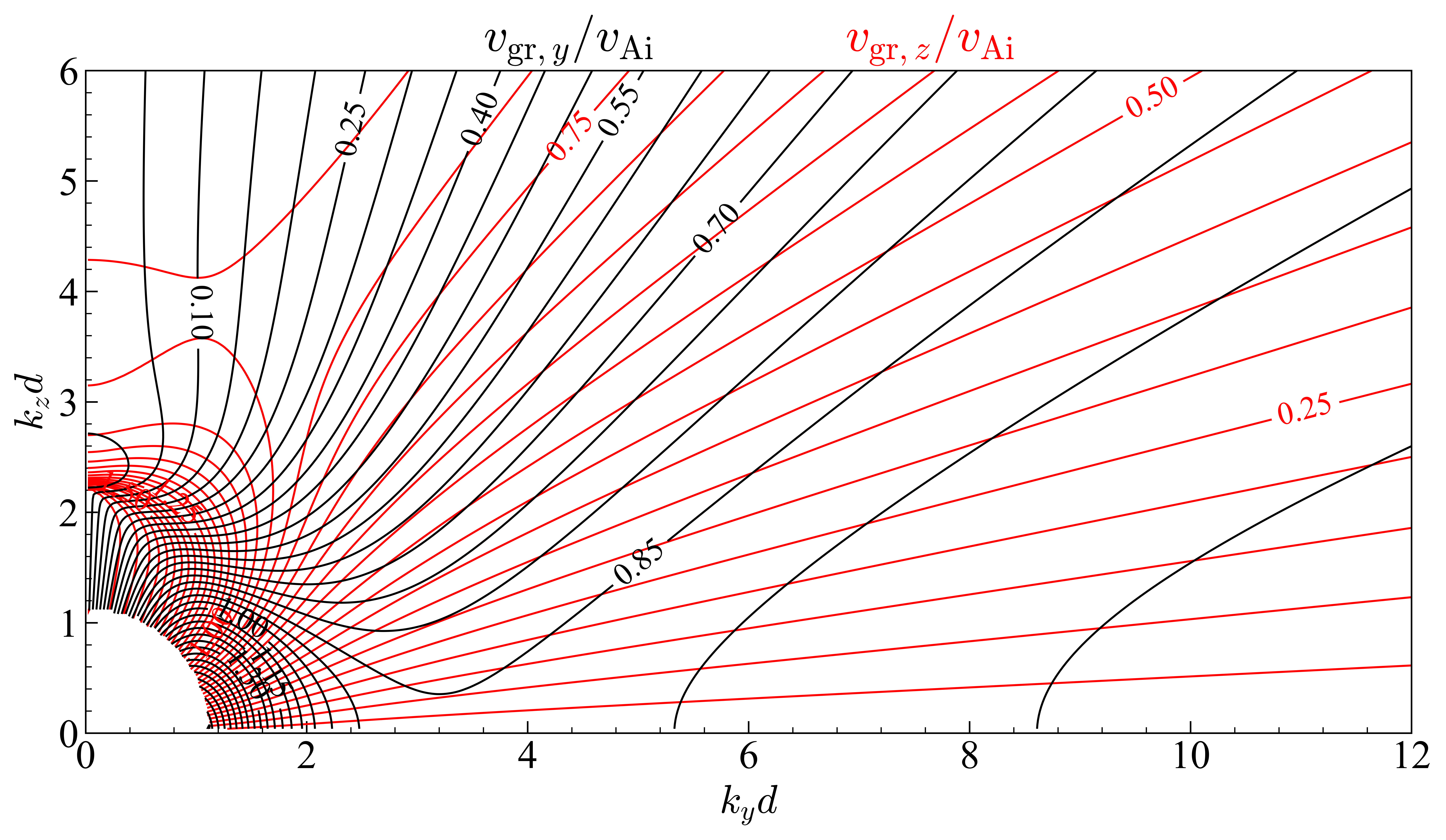}
\caption{
Similar to Fig.~\ref{fig_fund_vgr} but for 
	the first transverse overtone.
 }
\label{fig_overtone_vgr} 
\end{figure}

This subsection is devoted to the first transverse overtone. 
Figure~\ref{fig_overtone_vph} follows the format
    of Fig.~\ref{fig_fund_vph} to present the $[k_y, k_z]$-dependencies
    of the phase speed $\vph$ and the quantity $\myni$.
Note that trapped 3D modes are forbidden within the circle 
    $d\sqrt{k_y^2 + k_z^2} = k_z^{(1/2)}d = (\pi/2)/\sqrt{\rhoi/\rhoe -1}$,
    which reads $1.11$ given a $\rhoi/\rhoe=3$.
Note also that trapped 2D modes ($k_y=0$) are further prohibited
    for $k_z^{(1/2)} < k < k_z^{(1)} = 2k_z^{(1/2)}$.
When allowed, trapped modes belong exclusively to the body type ($\myni^2 >0$), 
    meaning that $\myni$ makes sense.
Figure~\ref{fig_overtone_vgr} presents, in a form identical to Fig~\ref{fig_fund_vgr},
    the distributions in the $k_y-k_z$ plane of 
    the $y$ component ($\vgy$) and $z$-component ($\vgz$)
    of the group velocity. 

It proves convenient to make analytical progress in various limiting cases 
    to help understand Figs.~\ref{fig_overtone_vph} and \ref{fig_overtone_vgr}.
First consider highly oblique propagation $k_y^2/k_z^2 \gg 1$. 
One readily sees from the DR (Eq.~\eqref{eq_DR_cot}) 
    that $\myni\to \pi^{-}$ when $k_y^2/k_z^2 \to \infty$, 
    meaning that the leading order solution $\omghat$ reads
    \begin{equation}
        \label{eq_overtone_largeKY_omghat}
                 \omghat(k_y, k_z) 
        \approx  \dfrac{\vAi}{d}
                 \sqrt{\pi^2 + (k_y d)^2 + (k_z d)^2}.
    \end{equation}
It then follows that $\vph=\omega/\sqrt{k_y^2+k_z^2}$ is approximately
    proportional to $\sqrt{1+\pi^2/[(k_y d)^2+(k_z d)^2]}$, meaning that
    the contours of $\vph$ eventually become a series of circles centered
    at the origin on the $k_y-k_z$ plane.
This is what one sees from Fig.~\ref{fig_overtone_vph}.
Furthermore, Equation~\eqref{eq_overtone_largeKY_omghat} indicates that
    \begin{equation}
        \dfrac{\vgy}{\vAi} 
        \approx \dfrac{1}{\sqrt{1+[\pi^2+(k_z d)^2]/(k_y d)^2}}.
    \end{equation}
At a given $k_z$, this means that $\vgy$ approaches the internal \Alf\ speed $\vAi$
    as a monotonically increasing function of $k_y$, in close agreement 
    with Fig.~\ref{fig_overtone_vgr}.
Likewise, one sees from Eq.~\eqref{eq_overtone_largeKY_omghat} that
    \begin{equation}
        \dfrac{\vgz}{\vAi} 
        \approx \dfrac{1}{\sqrt{1+[\pi^2+(k_y d)^2]/(k_z d)^2}},
    \end{equation} 
    which further simplifies to $\vgz \propto 1/\sqrt{1+(k_y/k_z)^2}$
    when $(k_y d)^2 \gg \pi^2$.
This means that the contours of $\vgz$ eventually become a series of rays
    radially directed from the origin on the $k_y-k_z$ plane, thereby
    explaining the red contours for $k_y^2/k_z^2 \gg 1$ in Fig.~\ref{fig_overtone_vgr}.
For completeness, we note that an improved version of Eq.~\eqref{eq_overtone_largeKY_omghat}  
    can be readily derived by writing 
    \begin{equation}
        \label{eq_overtone_largeKY_tmp}
        \omega = \omghat \left(1 + C_3 \epsilon^3 + C_4 \epsilon^4 + \cdots\right),
    \end{equation}
    with $\epsilon=k_z \vAi/\omghat$ seen as a small parameter.
Expanding both sides of Eq.~\eqref{eq_DR_cot}, 
    one readily verifies that the $\epsilon$-corrections start with third order 
    in the parentheses of Eq.~\eqref{eq_overtone_largeKY_tmp}.
An approximate solution that corrects $\omghat$ to forth order reads     
    \begin{equation}
    \label{eq_overtone_largeKY_Q3Q4}
        \omega(k_y, k_z)
      \approx 
        \omghat
        \left[
         1 - \dfrac{Q_3}{(\omghat d/\vAi)^3} + \dfrac{Q_4}{(\omghat d/\vAi)^4}
        \right],
    \end{equation}
    where
    \begin{subequations}
        \begin{align}
          & Q_3 = \dfrac{\pi^2}{\sqrt{(\rhoi/\rhoe)(\rhoi/\rhoe-1)}}, \\
          & Q_4 = \dfrac{3}{2}
                  \dfrac{\pi^2}{(\rhoi/\rhoe)(\rhoi/\rhoe-1)}.
        \end{align}
    \end{subequations}

Now consider ``small'' values of $k_y d$.
Some subtlety arises regarding what ``small'' means
    given the zone where trapped modes are forbidden.
Three cases need to be discriminated in terms of $k_z$. 
We start with Case I where $0 < k_z < k_{z}^{(1/2)}$.
As indicated by Fig~\ref{fig_charSols}a, the dispersion curve in this case
    starts from the intersection ($P$) between the curves $\me=0$ and $\myni=\pi/2$
    therein.
See $k_z$ as given, and let the coordinates of $P$ be denoted by 
    $[k_y^{\rm cut}, \omega^{\rm cut}]$.
One finds with the definitions of $\me$ and $\myni$ 
    (Eqs.~\eqref{eq_def_mie} and \eqref{eq_def_ni}) that
    \begin{subequations}
    \label{eq_overtone_case1_kyomgcut}
    \begin{align}
        & k_y^{\rm cut} = \sqrt{[k_z^{(1/2)}]^2 - k_z^2}, \\
        &    \omega^{\rm cut} 
          =  \dfrac{\pi/2}{\sqrt{1-\rhoe/\rhoi}}
             \dfrac{\vAi}{d}
          = k_z^{(1/2)} \vAe.
    \end{align}
    \end{subequations}
Evidently, $k_y^{\rm cut}$ is some critical out-of-plane wavenumber, 
    only beyond which trapped 3D modes are allowed for a given $k_z$.
Defining a small parameter $0<\epsilon \ll 1$ as
    \begin{equation}
    \label{eq_overtone_case1_DefEpsilon}
        k_y = k_y^{\rm cut}(1+\epsilon),    
        \quad \text{i.e., }
        \epsilon = k_y/k_y^{\rm cut} -1,
    \end{equation}
    we proceed to examine how trapped modes behave in the immediate vicinity of the forbidden zone by writing
    \begin{equation}
    \label{eq_overtone_case1_omg}  
        \omega(k_y, k_z)
      = 
        \omega^{\rm cut} 
        \left[
         1 + Q_1 \epsilon + Q_2 \epsilon^2 + \cdots
        \right]. 
    \end{equation}
Some tedious algebra yields that
    \begin{subequations}
    \label{eq_overtone_case1_Q1Q2}
    \begin{align}
         Q_1 &= 1-\dfrac{k_z^2}{(k_z^{(1/2)})^2},     \\
         Q_2 &= \dfrac{1}{2(k_z^{(1/2)} d)^2}
                \left[1
                    -\dfrac{k_z^2}{(k_z^{(1/2)})^2}
                \right]        \nonumber \\
             &\quad
                 \left\{
                   (k_z d)^2
                  - \left(\dfrac{\pi}{2}\right)^4
                    \dfrac{[1           -k_z^2/(k_z^{(1/2)})^2]^3}
                          {[\rhoi/\rhoe -k_z^2/(k_z^{(1/2)})^2]^2}
                \right\}. 
    \end{align}
    \end{subequations}
Evidently, both $Q_1$ and $Q_2$ approach zero when $k_z \to k_z^{(1/2)}$, 
    meaning that the expansion \eqref{eq_overtone_case1_omg} 
    is of little use therein.  
Furthermore, $Q_2$ is a bit intricate in that it switches from negative 
    to positive values when $k_z$ increases through some critical value.
Regardless, both $Q_1$ and $Q_2$ are 
    involved in the approximate expression for $\me$,
    \begin{equation}
    \label{eq_overtone_case1_me}    
        \me 
    \approx 
        \epsilon 
        \sqrt{(k_y^{\rm cut})^2 + (k_z^{(1/2)})^2 (Q_1^2 + 2 Q_2)}, 
    \end{equation}
    thereby providing a direct measure of the trapping capability of the slab
    when $k_y$ deviates slightly from $k_y^{\rm cut}$.

Equation~\eqref{eq_overtone_case1_omg} 
    proves largely adequate for explaining
    Figs.~\ref{fig_overtone_vph} and \ref{fig_overtone_vgr},
    provided that $k_z$ is not that close to $k_z^{(1/2)}$. 
Take the $\myni$ distribution in Fig.~\ref{fig_overtone_vph} for instance.
One finds from Eq.~\eqref{eq_overtone_case1_omg} that
    \begin{equation}
    \label{eq_overtone_case1_ni}
            \myni 
    \approx \dfrac{\pi}{2}
            \left[1+ \dfrac{(k_y^{\rm cut})^2}{(k_z^{(1/2)})^2} \epsilon\right],
    \end{equation}
    suggesting that $\myni-\pi/2$ scales largely as $\propto \epsilon$
    when $k_y$ deviates slightly from $k_y^{\rm cut}$ for a given $k_z$.
One further finds that
    \begin{equation}
    \label{eq_overtone_case1_vph}
               \vph 
      \approx  \vAe
               \left[1+\dfrac{Q_1^2 + 2 Q_2 - Q_1}{2} \epsilon^2
               \right].
    \end{equation}    
The phase speed $\vph$ therefore tends to decrease from 
    the external \Alf\ speed at a rather slow rate
    ($\propto \epsilon^2$).
Now consider Fig.~\ref{fig_overtone_vgr}.
It follows from Eq.~\eqref{eq_overtone_case1_omg} that 
    \begin{equation}
    \label{eq_overtone_case1_vgy}
              \vgy 
     \approx  \dfrac{\omega^{\rm cut}}{k_y^{\rm cut}}
              (Q_1 + 2 Q_2 \epsilon)
          =   \vAe \sqrt{1-\dfrac{k_z^2}{(k_z^{(1/2)})^2}}
             +2Q_2 \dfrac{\omega^{\rm cut}}{k_y^{\rm cut}}\epsilon. 
    \end{equation}
It immediately follows that $\vgy$ decreases from $\vAe$ toward zero
    when $k_z$ increases from zero toward $k_z^{(1/2)}$ along
    the outer edge of the forbidden zone. 
Furthermore, that $Q_2$ changes sign  
    means that $\vgy$ decreases (increases) from the edge value 
    when $k_y$ increases from $k_y^{\rm cut}$
    for a given $k_z$ that is below (above) some critical value. 
Equation~\eqref{eq_overtone_case1_omg} can also be used to derive 
    an approximate expression for $\vgz$.
This derivation is somehow complicated by the $k_z$-dependencies
    of $k_y^{\rm cut}$ and $\epsilon$
    (see Eqs.~\eqref{eq_overtone_case1_kyomgcut} and \eqref{eq_overtone_case1_DefEpsilon}).
Some algebra yields that
    \begin{equation}
    \label{eq_overtone_case1_vgz}    
              \vgz 
      \approx \vAe \dfrac{k_z}{k_z^{(1/2)}} 
              \left[
                  1+\left(
                       -1 + 2Q_2 \dfrac{(k_z^{(1/2)})^2}{(k_y^{\rm cut})^2}
                    \right)\epsilon
              \right] . 
    \end{equation}
Taking $\epsilon\to 0$, one deduces from
    Eq.~\eqref{eq_overtone_case1_vgz} 
    that $\vgz$ varies from zero toward $\vAe$ 
    when $k_z$ is surveyed along the edge of the forbidden zone
    from zero toward $k_z^{(1/2)}$.
The $\epsilon$-correction, one the other hand, immediately
    suggests that $\vgz$ decreases from its edge value
    when $k_y$ increases from $k_y^{\rm cut}$ for a $k_z$
    ensuring $Q_2 <0$.
This is in agreement with Fig.~\ref{fig_overtone_vgr}.
Somehow puzzling is that  
    the same $k_y$-dependence actually persists for those $k_z$
    that yield positive $Q_2$.
This can be addressed by expanding
    the $Q_2$-term in Eq.~\eqref{eq_overtone_case1_vgz}
    with the aid of Eq.~\eqref{eq_overtone_case1_Q1Q2},
    the result being
    \begin{eqnarray}
      &&  2Q_2 \dfrac{(k_z^{(1/2)})^2}{(k_y^{\rm cut})^2} \nonumber \\ 
      &=& \dfrac{1}{(k_z^{(1/2)} d)^2}
                 \left\{
                   (k_z d)^2
                  - \left(\dfrac{\pi}{2}\right)^4
                    \dfrac{[1           -k_z^2/(k_z^{(1/2)})^2]^3}
                          {[\rhoi/\rhoe -k_z^2/(k_z^{(1/2)})^2]^2}
                \right\}. 
    \end{eqnarray}
Evidently, this $Q_2$-term is consistently smaller than unity.
It then follows that the coefficient of $\epsilon$
    in Eq.~\eqref{eq_overtone_case1_vgz}
    is always negative.

We now move on to Case II where $k_{z}^{(1/2)} < k_z < k_{z}^{(1)}$.
This case is somehow peculiar in that there exist
    no trapped 2D modes ($k_y =0$),
    whereas trapped 3D modes suddenly arise 
    when $k_y$ becomes finite, regardless of how small $k_y d$ is.
One readily verifies that $\me \to 0$ when $k_y \to 0$, meaning that 
    the leading-order solution to Eq.~\eqref{eq_DR_cot} reads
    $\omega \approx k_z\vAe$.
Defining $\epsilon = k_y^2/k_z^2 \ll 1$, 
    we look for improved approximate solutions in the form
    $\omega = k_z\vAe (1+ Q_1 \epsilon - Q_2 \epsilon^2 +\cdots)$, or equivalently
    \begin{equation}
    \label{eq_overtone_case2_omg}
        \omega(k_y, k_z)
      \approx 
        k_z \vAe 
        \left[
         1 + Q_1 \dfrac{k_y^2}{k_z^2} - Q_2 \dfrac{k_y^4}{k_z^4} + \cdots
        \right].
    \end{equation}
Some algebra yields that 
    \begin{equation}
    \label{eq_overtone_case2_Q1Q2}
        Q_1 = \dfrac{1}{2}, \qquad
        Q_2 = \dfrac{1}{2}
              \left[
                 \dfrac{1}{4}
                +\dfrac{\cot^2(k_z d \sqrt{\rhoi/\rhoe-1})}
                       {\rhoi/\rhoe-1} 
              \right].
    \end{equation}
The reason for us to retain the $Q_2$ term is that
    it is involved in $\me^2$,
    which approximates to
    \begin{equation*}
        \me^2 \approx -k_z^2 (Q_1^2 - 2 Q_2) \epsilon^2.
    \end{equation*}
More specifically, one finds that    
    \begin{equation}
    \label{eq_overtone_case2_me}
        \me \approx \dfrac{\cot(k_z d \sqrt{\rhoi/\rhoe-1})}
                          {\sqrt{\rhoi/\rhoe-1}}
                    \dfrac{k_y^2}{k_z}.      
    \end{equation}
For a fixed $k_z$, 
    the slab therefore possesses some poorer trapping capability     
    ($\me \propto k_y^2$) than in Case I where $\me \propto (k_y - k_y^{(\rm cut)})$ (Eq.~\eqref{eq_overtone_case1_me}). 
More importantly, that $\me$ possesses a continuous $k_y$-dependence means that
    the sudden appearance of trapped 3D modes for $k_y>0$
    does not result in any abrupt change in the 
    the temporal response of the slab to impulsive, localized, 3D drivers.

Equation~\eqref{eq_overtone_case2_omg} 
    proves adequate for explaining some key features of
    Figs.~\ref{fig_overtone_vph} and \ref{fig_overtone_vgr}.
One deduces from Eq.~\eqref{eq_overtone_case2_omg} that
    \begin{equation}
    \label{eq_overtone_case2_ni}
    \myni \approx \sqrt{\rhoi/\rhoe-1} 
                  \sqrt{k_y^2 + k_z^2}.
    \end{equation}
When $k_y \to 0$, the quantity $\myni$ therefore increases monotonically
    from $\pi/2$ to $\pi$ as $k_z$ increases from $k_z^{(1/2)}$ to $k_z^{(1)}$.
The contours of $\myni$ for small $k_y$, on the other hand, are expected to be
    concentric circles as indeed seen in Fig.~\ref{fig_overtone_vph}.
One further deduces that
    \begin{equation}
    \label{eq_overtone_case2_vph}
    \vph \approx \vAe 
                 \left[1-
                         \dfrac{1}{2}
                         \dfrac{\cot^2(k_z d \sqrt{\rhoi/\rhoe-1})}
                               {(\rhoi/\rhoe-1)} 
                        \dfrac{k_y^4}{k_z^4}
                 \right],
    \end{equation}
    meaning that $\vph$ decreases from $\vAe$  
    at a very slow rate ($\propto k_y^4$) when $k_y$ increases
    for a given $k_z$. 
Now move on to Fig.~\ref{fig_overtone_vgr}, to explain which
    it suffices to retain only the $Q_1$ term in 
    Eq.~\eqref{eq_overtone_case2_omg}.
One finds that   
    \begin{equation}
    \label{eq_overtone_case2_vgy}
        \vgy \approx \vAe \dfrac{k_y}{k_z},
    \end{equation}
    thereby explaining why the $\vgy$ contours 
    are largely radially directed.
One further finds that 
    \begin{equation}
    \label{eq_overtone_case2_vgz}
            \vgz 
    \approx \vAe 
            \left(1-\dfrac{k_y^2}{2 k_z^2}
            \right),
    \end{equation}
    meaning that $\vgz$ for a given $k_z$ tends
    to decrease rather slowly with $k_y$.
Equation~\eqref{eq_overtone_case2_vgz} also explains why
    the $\vgz$ contours are largely radially aligned
    when $k_z$ is not that close to $k_z^{(1)}$.

We now address Case III where $k_z > k_z^{(1)}$.
It is textbook material that trapped 2D modes ($k_y = 0$) are now allowed,
    their phase speeds $\omgtwoD(k_z)/k_z$ decreasing monotonically
    from $\vAe$ to $\vAi$ when $k_z$ increases from $k_z^{(1)}$
    to large values
    \citep[e.g.,][Chapter~5]{2019CUP_Roberts}.
One further finds that 
	\begin{equation}
     \label{eq_overtone_ky0_kzsmall}
	           \omgtwoD(k_z) 
          \approx k_z \vAe 
                  \left[
                    1-\dfrac{(\rhoi/\rhoe-1)\pi^2}{2}
                      \left(\dfrac{k_z}{k_z^{(1)}}-1\right)^2
                  \right] 
    \end{equation}
    when $k_z$ exceeds ${k_z^{(1)}}$ only marginally.
A similar expression is available for trapped 2D sausage modes
    \citep[][Eq.~(22)]{2018ApJ...855...53L}, 
    and Eq.~\eqref{eq_overtone_ky0_kzsmall} can be derived with
    the same procedure therein. 
Likewise, one finds that
	\begin{equation}
     \label{eq_overtone_ky0_kzlarge}
	           \omgtwoD(k_z) 
          \approx k_z \vAi 
                  \sqrt{1+(3\pi/2)^2 (k_z d)^{-2}},
    \quad 
         k_z d \gg 1,
    \end{equation}
    by specializing Eq.~(16) in \citet{2013ApJ...767..169L}
    to the pressureless limit.
Note that the quantity $\myni=\sqrt{(\omgtwoD/\vAi)^2-k_z^2}$ for 
    trapped 2D overtones
    behaves similarly to transverse fundamentals in that
    $\myni$ also increases monotonically with $k_z$
    (see Fig.~\ref{fig_overtone_vph}).
Equations~\eqref{eq_overtone_ky0_kzsmall} and 
    \eqref{eq_overtone_ky0_kzlarge} 
    help explicitly show that $\myni d\to \pi$ when $k_z \to k_z^{(1)}$
    and $\myni d \to 3\pi/2$ when $k_z d \to \infty$.

\begin{figure}
\centering
\includegraphics[width=.99\columnwidth]{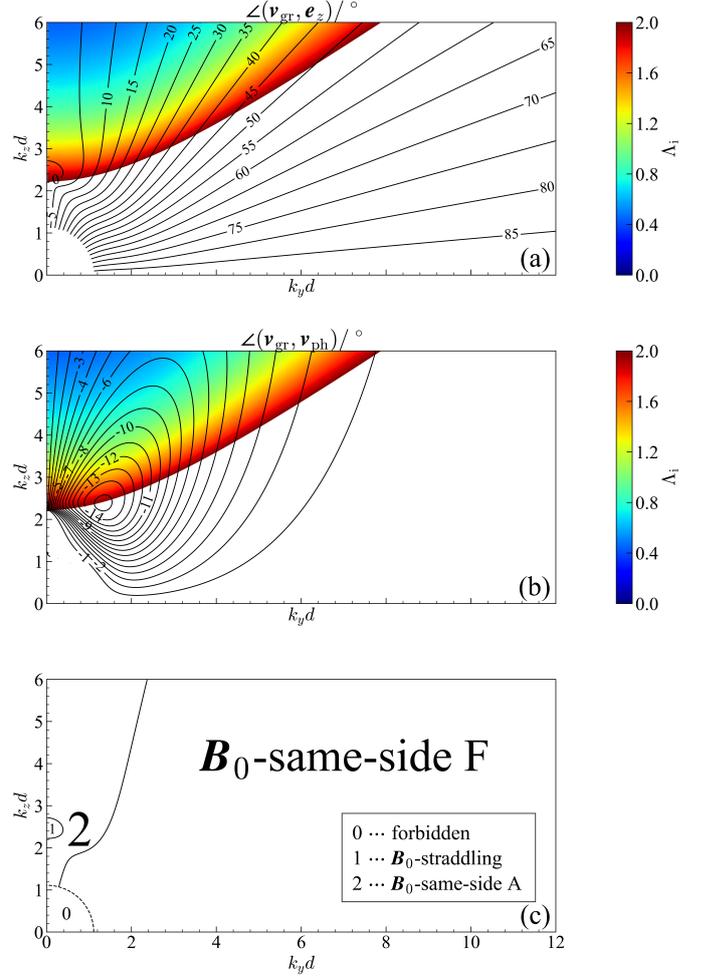}
\caption{
Similar to Fig.~\ref{fig_fund_angles} but for 
	the first transverse overtone.
Only a limited range is plotted for the quantity $\Lambdai$
    because $\Lambdai$ takes up too diverse a range.
Panel c overviews how the overtone behaves on the $k_y-k_z$ plane.
Trapped 3D modes are forbidden where
    $\sqrt{k_y^2 + k_z^2} \le k_z^{(1/2)}$.
When allowed, the overtone is further classified into three regimes
    based on the angle information.
See text for more details. 
}
\label{fig_overtone_angles} 
\end{figure}

Now consider what happens when $0<k_y^2/k_z^2 \ll 1$.
One readily verifies that 
    Eq.~\eqref{eq_fund_smallKY} holds for transverse overtones as well, 
    as long as 
    the quantities $\omgtwoD$ and $\metwoD$ are 
    interpreted appropriately.
We rewrite Eq.~\eqref{eq_fund_smallKY} as
    \begin{equation}
    \label{eq_overtone_case3_omg}    
            \omega(k_y, k_z)
	\approx \omgtwoD(k_z) 
			[1+ \mathcal{C}(k_z) (k_y d)^2]     
    \end{equation}
    for the ease of description.
It follows from Eqs.\eqref{eq_overtone_ky0_kzsmall} and 
    \eqref{eq_overtone_ky0_kzlarge}
    that $\metwoD =\sqrt{k_z^2 - (\omgtwoD)^2/\vAe^2} \propto (k_z/k_z^{(1)}-1)$ when $k_z$ barely exceeds $k_z^{(1)}$,
    while $\metwoD \propto k_z$ when $k_z d \gg 1$.
The quantity $\metwoD$ itself turns out to be a monotonically increasing
    function of $k_z$.
Equation~\eqref{eq_fund_smallKY} then indicates that   
    there must exist some critical axial wavenumber, $k_z^{\rm 2D,c}$,
    such that $\mathcal{C} < 0$ ($\mathcal{C} > 0$) 
    when $k_z^{(1)} <k_z < k_z^{\rm 2D,c}$ ($k_z > k_z^{\rm 2D,c}$).
One therefore deduces from Eq.~\eqref{eq_overtone_case3_omg} that
    $\vgy=\partial\omega/\partial k_y$ varies from zero when $k_y\to 0$
    toward negative (positive) values 
    for those values of $k_z$ that ensure $\mathcal{C} < 0$ ($\mathcal{C} > 0$).
This is what one sees from Fig.~\ref{fig_overtone_vgr}, 
    and the $\vgy=0$ contour therein actually identifies 
    $k_z^{\rm 2D,c}d \approx 2.74$ for the chosen $\rhoi/\rhoe=3$.
Restrict ourselves to $k_z^{(1)} <k_z < k_z^{\rm 2D,c}$.
Figure~\ref{fig_overtone_vgr} shows the subtle feature that
    $\vgz$ tends to decrease (increase) from its edge value ($k_z \to 0$)
    when $k_y$ increases for relatively small (large) values of $k_z$.
This can be adequately addressed by taking the $k_z$-derivative 
    of Eq.~\eqref{eq_overtone_case3_omg}, 
    even though the cumbersome $\mathd \mathcal{C}/\mathd k_z$
    term somehow complicates the matter.

Figure~\ref{fig_overtone_angles} applies the same format
    of Fig.~\ref{fig_fund_angles}
    and the associated classification scheme to the first
    transverse overtone. 
Note that the filled contours display 
    only a limited range for the force ratio $\Lambdai$, 
    despite that $\Lambdai$ can be evaluated wherever
    trapped modes are permitted.
Consider trapped 2D modes first ($k_y=0$), 
    which are recalled to arise for $k_z > k_z^{(1)}$.
One readily finds from Eqs.~\eqref{eq_overtone_ky0_kzsmall}
    and \eqref{eq_overtone_ky0_kzlarge}
    that $\Lambdai\to \rhoi/\rhoe-1$ when $k_z$ approaches $k_z^{(1)}$
        from above,
        and $\Lambdai\to 0$ when $k_zd \to \infty$.
For highly oblique propagation ($k_y^2/k_z^2 \gg 1$),
    however, Equation~\eqref{eq_overtone_largeKY_Q3Q4} suggests that
    $\Lambdai=\omega^2/k_z^2\vAi^2-1$ approximates to 
    $[\pi^2+(k_y d)^2)]/(k_z d)^2$ and may therefore take
    extremely large values.

Figure~\ref{fig_overtone_angles}c overviews how the first overtone
    behaves on the $k_y-k_z$ plane.
The disc $\sqrt{k_y^2+k_z^2}$, labeled $0$, is recalled to be where
    trapped 3D modes are forbidden.
Our classification scheme is then applied,
    and all the three subgroups are now relevant.
 {The ``$\vec{B}_0$-straddling''} subgroup (with $\vgy < 0$, labeled $1$) remains associated with
    force ratios $\Lambdai \gtrsim 1$, 
    despite that this subgroup is restricted to 
    a substantially smaller portion when compared with the fundamental. 
 {The ``$\vec{B}_0$-same-side F''} subgroup corresponds to the vast portion to the right
    of the $|\angle(\vecvgr, \uvec{z})| = 15^\circ$ contour.
Our additional criterion $|\angle(\vecvgr, \vecvph)|<15^\circ$ does not 
    provide further constraints per se.
In fact, Figure~\ref{fig_overtone_angles}b 
    indicates a small $\angle(\vecvgr, \vecvph)$ almost everywhere, which 
    is particularly true for oblique propagation ($k_y^2/k_z^2\gg 1$).
This feature, in conjunction with the overall tendency for 
    $\Lambdai$ to be (relatively) large, means that  {the ``$\vec{B}_0$-same-side F''} subgroup 
    somehow closely resembles what happens in unbounded uniform media.
However,  {the ``$\vec{B}_0$-same-side A''} 
    subgroup deviates considerably from its counterpart in an
    unbounded uniform medium, in which
    case one recalls a $\Lambda \lesssim 0.07$.
The force ratio $\Lambdai$ for  {the ``$\vec{B}_0$-same-side A''} 
    subgroup here, in contrast, 
    may readily exceed $\rhoi/\rhoe-1$.
 {
Note that $|\angle(\vecvgr, \uvec{z})|$ evaluates to $45^\circ$ 
    when $\Lambda=1$ in an unbounded uniform medium
    (see Eq.~\eqref{eq_Lambda_uni2}). 
Our specific threshold of $15^\circ$ 
    for mode categorization is therefore only for illustrative purposes.
A larger threshold (up to, say, $45^\circ$) does not impact
    the categorization of the transverse fundamental,
    given that $|\angle(\vecvgr, \uvec{z})|$ does not exceed 
    $\sim 11^\circ$ (Fig.~\ref{fig_fund_angles}).
Nonetheless, a larger threshold will broaden 
    the $k_y-k_z$ portion where the first overtone
    qualifies as ``$\vec{B}_0$-same-side A''
    in Fig.~\ref{fig_overtone_angles}c.
The rest of the ``$\vec{B}_0$-same-side'' portion always qualifies
    as ``$\vec{B}_0$-same-side F'' modes; 
    $|\angle(\vecvgr, \vecvph)|$ is consistently smaller than $\sim 15^\circ$
    and therefore not a discriminating factor.
Importantly, the threshold angle is not involved hereafter;
    only the distinction between ``$\vec{B}_0$-straddling''
    and ``$\vec{B}_0$-same-side'' is employed. 
}

\section{Implications for impulsively excited 3D kink motions}
\label{sec_impulsive}
How does the above-presented mode analysis connect to the spatio-temporal
    evolution of a density-enhanced slab in response to localized initial exciters?
We start by noting that the problem at hand is an initial value problem
    (IVP) governed by the time-dependent MHD equations
    \eqref{eq_linMHD_momen} and \eqref{eq_linMHD_Farad}
    over the infinite $[x,y,z]$-volume.
Conceptually, the nominal boundaries $x, y, z\to \pm \infty$ 
    need to be such that no boundary conditions (BCs) are specified 
    or equivalently that the dependent variables do not diverge
    \citep[see e.g.,][and references therein]{2024A&A...692A.259G}.
The most unambiguous way to specify initial exciters, on the other hand,
    is to perturb the velocities only 
    ($\vec{v}_{1}(x,y,z, t=0)\ne 0$, $\vec{B}_{1}(x,y,z, t=0) = 0$;
    see Sect.~6.1.2 in \citealt{2019CUP_goedbloed_keppens_poedts}).
We proceed with the following assumptions for definitiveness.
Firstly, the only non-vanishing 
    independent quantity at $t=0$ is $v_{1x}$.
Secondly, $v_{1x}(x,y,z, t=0)$ is even in $x$ to ensure kink parity,
    and is even in both $y$ and $z$ as well.
Thirdly, the implementation of $v_{1x}(x,y,z, t=0)$
    is such that one needs only to consider the range of $[k_y, k_z]$
    examined in Figs.~\ref{fig_fund_vgr} and \ref{fig_overtone_vgr}. 

Quantitative progress can be made by presenting 
    the formal solution to the IVP
    \citep{2023MNRAS.518L..57L},
    \begin{eqnarray}
     &&   v_{1x}(x,y,z,t)
     = \int_{-\infty}^{\infty} \mathd k_y
        \int_{-\infty}^{\infty} \mathd k_z    \nonumber\\
     && \quad 
        \left\{
        \sum_j \left[\mathcal{F}_j(x; k_y, k_z)
                     \Exp{\imath (\omega_j t - k_y y - k_z z)}
               \right]
        +\text{improper}       
        \right\}.        \label{eq_3DIVP_formalSol}
    \end{eqnarray}
Equation~\eqref{eq_3DIVP_formalSol} is so written by following 
    the approach of Fourier integrals \citep[e.g.,][]{2014ApJ...789...48O,2015ApJ...806...56O} 
        or equivalently the idea of decomposing the time-dependent wave fields
        into the eigenmodes of the MHD force operator
        \citep[e.g.,][]{2022ApJ...928...33L,2023ApJ...943...91W}.
We adopt the nomenclature of the latter.
The decomposition in the $y$- or $z$-direction is self-evident, 
    and one then ends up with an eigenvalue problem with
    $x$ being the only independent variable and $[k_y, k_z]$ serving
    as parameters. 
At any given pair $[k_y, k_z]$, the eigenspectrum of the MHD force operator
    for our slab equilibrium comprises two subspectra.
The point subspectrum is populated by a finite number of 
    eigensolutions, which are labeled by $j$ and are equivalent
    to our trapped modes ($\me^2>0$).
By construction, $\omega_j = \omega_j(k_y, k_z)$ is dictated by
    the DR.
The other subspectrum (with $\me^2<0$), called ``improper continuum'',
    comprises those eigensolutions whose eigenfrequencies continuously
    cover the range from some critical frequency out to infinity.
This ``improper continuum'' arises in our context as a result of 
    the physical requirement
    that no BCs should be specified at $x\to \pm \infty$.
Regardless, the contribution from the continuum eigenmodes
    (labeled ``improper'' in Eq.~\eqref{eq_3DIVP_formalSol}) attenuates rapidly with time and hence are not of further interest here. 
    
What survives at large times is the contribution from
    trapped modes. 
We focus on the slab axis $x=0$, supposing further that
    only the transverse fundamental and its first overtone
    are relevant.
These simplifications suffice to illustrate 
    both the usefulness of and intricacies in the method of stationary phase
    (MSP) that we adopt to digest the morphological features
    of wave propagation. 
Suppose that the MSP applies to some $[y, z, t]$ or more appropriately
    to some $[y/t, z/t]$ with $[y, z, t]$ seen as given.
Equation~\eqref{eq_3DIVP_formalSol}, while nominally involving 
    all $[k_y, k_z]$, is actually dominated only by
    a series of narrow regions (or wavepackets in physical terms)
    on the $k_y-k_z$ plane.
Let wavepackets (WPs) be numbered by $n$
    and represented by the central wavenumbers $[K_{n, y}, K_{n, z}]$.
It follows from the general theory of the MSP that
    \citep[e.g.,][Chapter~11]{1974Whitham} 
    \begin{equation}
    \label{eq_3DIVP_MSPv1x}
           v_{1x}(x=0, y, z, t) 
      \sim t^{-1} 
           \sum_n  
               \left[\mathcal{G}_n(K_{n, y}, K_{n, z})
                     \Exp{\imath (\omega_n t - K_{n, y} y - K_{n, z} z)}
               \right],
    \end{equation}
    where $\omega_n = \omega_n(K_{n, y}, K_{n, z})$ 
    is given by the DR. 
Involved in the summation is any WP that solves the kinematic equation
    \begin{equation}
    \label{eq_3DIVP_MSPkinematic}
        \vgy(K_{n, y}, K_{n, z}) = y/t, \quad
        \vgz(K_{n, y}, K_{n, z}) = z/t.
    \end{equation}
It suffices to consider only the first quadrant ($y, z>0$) given 
    the assumed $y$- and $z$-symmetries of the 
    exciter $v_{1x}(x,y,z,t=0)$. 
Some subtlety arises now in that so far our mode analysis addresses
    only the first quadrant in the wavenumber plane 
    ($k_y>0, k_z>0$), with $\vgy<0$ for our  {``$\vec{B}_0$-straddling'' modes}.
WPs associated with  {these ``$\vec{B}_0$-straddling'' modes} remain possible to show up
    for $[y>0, z>0]$.
Suppose that some $[K'_{n, y}>0, K'_{n, z}>0]$ satisfies 
    \begin{equation*}
        \vgy(K'_{n, y}, K'_{n, z}) = -y/t, \quad
        \vgz(K'_{n, y}, K'_{n, z}) = z/t.
    \end{equation*}
The symmetry discussions on Eq.~\eqref{eq_omgFormal}
    then dictate that WP 
    $[K_{n, y}=-K'_{n, y}, K_{n, z}=K'_{n, z}]$ solves Eq.~\eqref{eq_3DIVP_MSPkinematic}.

\begin{figure}
\centering
\includegraphics[width=.99\columnwidth]{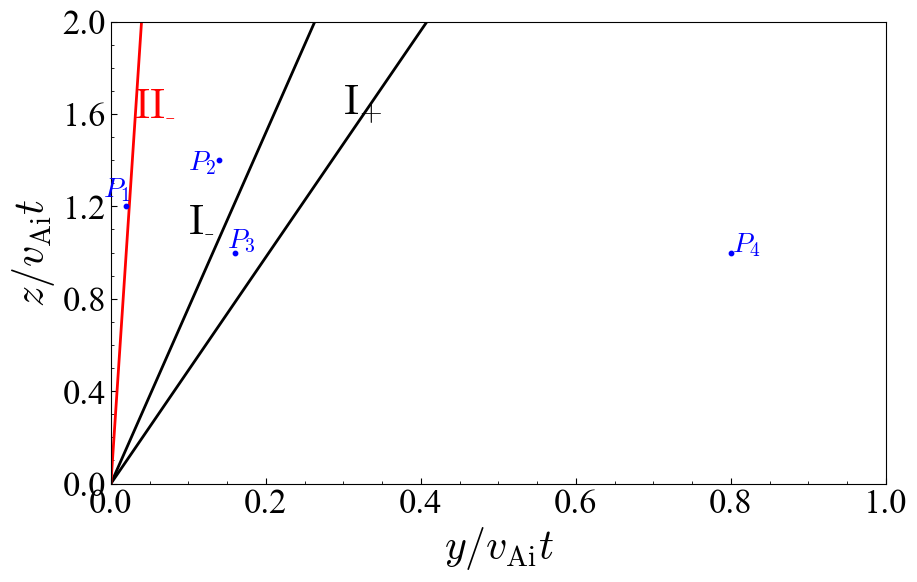}
\caption{Categorization of the $y/t-z/t$ plane in terms of large-time expectations
    for the morphological features of impulsive kink motions.
The bordering straight lines are determined by the limiting behavior of 
    the group velocities as derived from Figs.~\ref{fig_fund_vgr} and \ref{fig_overtone_vgr}.
Line ${\rm I}_{+}$ pertains to  {``$\vec{B}_0$-same-side'' fundamental modes}, 
    while
    line ${\rm I}_{-}$ (${\rm II}_{-}$) pertains to  {``$\vec{B}_0$-straddling'' modes} associated
    with the transverse fundamental (first overtone).
Four sectors result.
Different morphological features are expected for different sectors, 
    the difference being particularly pronounced when line ${\rm I}_{-}$ is crossed.
One representative point in each sector
    (marked $P_1$ to $P_4$) is selected for further quantitative analysis.
See text for more details.
}
\label{fig_3DIVP_yzplane} 
\end{figure}

Figure~\ref{fig_3DIVP_yzplane} collects the results of our mode analysis
    to show what WPs are expected in different portions
    of the $y/t-z/t$ plane.
The horizontal and vertical axes are not to the same scale,
    otherwise Fig.~\ref{fig_3DIVP_yzplane} will be difficult to present.
Different portions are discriminated by the different straight lines,
    whose slopes ($S>0$) are determined by Eq.~\eqref{eq_3DIVP_MSPkinematic}.
Specifically,
    \begin{itemize}
        \item the slope of line ${\rm II}_{-}$ reads $S_{\rm II-}=50.6$, the minimal $|\vgz/\vgy|$ that  {``$\vec{B}_0$-straddling'' overtone modes} may reach.
        \item the slope of line ${\rm I}_{-}$ reads $S_{\rm I-}=7.6$, the minimal $|\vgz/\vgy|$ that  {``$\vec{B}_0$-straddling'' fundamental modes} may reach.
        \item the slope of line ${\rm I}_{+}$ reads $S_{\rm I+}=4.9$, the minimal $\vgz/\vgy$ reachable by  {``$\vec{B}_0$-same-side'' fundamental modes}.
    \end{itemize}
Four sectors are eventually discriminated,
    with Eq.~\eqref{eq_3DIVP_MSPkinematic} determining the types of WPs
    that each sector allows.
One way to visualize this is that Fig.~\ref{fig_3DIVP_yzplane} is superimposed by
    all pairs of $[\vgy, \vgz]$ in Figs.~\ref{fig_fund_vgr} and \ref{fig_overtone_vgr},
    with the sign of $\vgy$ reversed for  {``$\vec{B}_0$-straddling'' modes}.
Overall, two clouds will show up, one for the transverse fundamental
    and the other for its first overtone. 
Wave signatures can be found only in these clouds, if one sees $t$ in 
    Eq.~\eqref{eq_3DIVP_MSPkinematic} as some given large time
    and therefore translates the $y/t-z/t$ plane 
    into the more intuitive $y-z$ plane.
We choose not to do this because the resulting figure is way too crowded.
Rather, we somehow arbitrarily choose one space-time point from the cloud(s)
    within each sector in Fig.~\ref{fig_3DIVP_yzplane}.
The coordinates $[y/t, z/t]$ of these points, labeled $P_1$ to $P_4$, are then
    inverted with Eq.~\eqref{eq_3DIVP_MSPkinematic} for 
    the allowed WPs. 
Table~\ref{tab_WPsolutions} then results, where the third and fourth columns
    present the central wavenumbers of the inverted WPs
    from the fundamental and first overtone, respectively.

Some key morphological features at some arbitrarily given large $t$ 
    can be predicted for wave propagation in the 
    first quadrant ($y,z>0$) of the $x=0$ plane.
We focus on the wavefronts (i.e., isophase curves) of $v_{1x}$, 
    specializing to the $v_{1x}(x=0, y, z, t)=0$ contours for the ease
    of description. 
We further assume that only one WP dominates
    the summation in Eq.~\eqref{eq_3DIVP_MSPv1x} for a given pair $[y/t, z/t]$, 
    even though Eq.~\eqref{eq_3DIVP_MSPkinematic} tends to admit
    multiple solutions in general.
Let $[\mathcal{K}_y, \mathcal{K}_z]$ denote this dominating WP.
The vector $\mathcal{K}_y \uvec{y} + \mathcal{K}_z \uvec{z}$ is then 
    aligned with the directions of both the local normal
    and the instantaneous propagation of 
    the instantaneous wavefront at the given $[y,z]$
    (see Eq.~\eqref{eq_3DIVP_MSPv1x}).
Now recall that $\vgz$ as a function of wavenumbers $[k_y, k_z]$
    is positive definite
    in Figs.~\ref{fig_fund_vgr} and \ref{fig_overtone_vgr}, 
    meaning that $\vgz$ for $[k_y>0, k_z>0]$ 
    is of the same sign as the $z$-component
    of the phase velocity 
    ($v_{{\rm ph}, z} = \omega k_z/(k_y^2+k_z^2)$, see Eq.~\eqref{eq_defVphVg_slab}). 
One readily verifies that the signs of 
    $\vgz$ and $v_{{\rm ph}, z}$ remain the same for all the eight
    combinations of the signs of $k_y$, $k_z$, and $\omega$, 
    given the symmetry properties of Eq.~\eqref{eq_omgFormal}.
Note that we have assumed, without loss of generality, that the initial exciter
    is localized around $[x=0,y=0, z=0]$.
It follows from causality considerations that any WP that can be inverted
    for a space-time point with $z>0$ must possess a positive $\vgz$, which 
    in turn must mean a positive $v_{{\rm ph},z}$ and hence a positive $\mathcal{K}_z$.
Consequently, the local normal of any wavefronts must be locally directed upwards. 
However, the possible relevance of  {``$\vec{B}_0$-straddling'' modes} substantially complicates
    the considerations regarding whether the local normal is directed
    leftward (i.e., toward $y=0$) or rightward (i.e., away from $y=0$).

The specific computations in Table~\ref{tab_WPsolutions} help 
    address whether leftward-directed wavefronts
    are allowed in each sector in Fig.~\ref{fig_3DIVP_yzplane}.
These expectations are collected as follows.
\begin{itemize}
    \item The sector bordered by the vertical axis and line ${\rm II}_{-}$ allows both leftward and rightward wavefronts. The WP dominating a space-time point along a leftward wavefront is necessarily associated with a  {``$\vec{B}_0$-straddling'' mode}, which may belong to
    the transverse fundamental or its first overtone. 
    \item The sector bordered by lines ${\rm II}_{-}$ and ${\rm I}_{-}$ allows both leftward and rightward wavefronts. The WP dominating a space-time point along a leftward wavefront derives from a  {``$\vec{B}_0$-straddling'' transverse fundamental mode}.
    \item The sector bordered by lines ${\rm I}_{-}$ and ${\rm I}_{+}$ allows only rightward wavefronts. The associated dominating WPs are  {exclusively ``$\vec{B}_0$-same-side''}, 
    and may be associated with the transverse fundamental or its first overtone.  
    \item The sector bordered by the horizontal axis and line ${\rm I}_{+}$ allows only
    rightward wavefronts. The associated dominating WPs  {are ``$\vec{B}_0$-same-side''}, and derive from the first overtone.
\end{itemize}
We choose to simplify our discussion by
    not further grouping  {``$\vec{B}_0$-same-side'' modes into the ``$\vec{B}_0$-same-side A'' and ``$\vec{B}_0$-same-side F'' ones}.    
This does not hinder our purpose for demonstrating the power of the MSP that enables
    one to understand the intricate interference patterns in impulsively excited waves.
Rather, these expectations are in close agreement with our previous
    results from a direct 3D numerical simulation \citep[][Fig.~3d in particular]{2023MNRAS.518L..57L}.

\begin{table}[tpb]
\caption{Wavepackets for a representative set of space-time points}
\label{tab_WPsolutions}
\centering
\begin{tabular}{l|l|l|l}
\hline
Point  & $\left[\dfrac{y}{\vAi t}, \dfrac{z}{\vAi t}\right]$ 
                                & \makecell[l]{fundamental WP(s) \\[0.6em]
                                              $[K_{n,y}d, K_{n,z}d]$}
                                & \makecell[l]{\small $1^{\rm st}$ overtone WP(s)    \\[0.6em]
                                              $[K_{n,y}d, K_{n,z}d]$} \\
\hline        
$P_1$  & $[0.02, 1.2]$          & \makecell[l]{$[-0.01,0.44]$ \\
                                               $[-1.26,0.66]$}           
                                & \makecell[l]{$[-0.089,2.31]$ \\
                                               $[-0.21,2.31]$ \\
                                               $[0.41,2.29]$} \\
\hline 
$P_2$  & $[0.14, 1.4]$          & \makecell[l]{$[-0.021,0.27]$ \\
                                               $[-0.22,0.26]$}
                                & $[0.52,2.12]$                   \\
\hline
$P_3$  & $[0.16, 1]$            & $[2.66,4.18]$      
                                & $[1.04,2.34]$       \\
\hline
$P_4$  & $[0.8, 1]$             & None                  
                                & $[1.32,1.31]$       \\
\hline
\end{tabular}
\end{table}

 {
Some remarks can be offered on the possible seismic applications of 
    our theoretical findings to the imaging observations
    of impulsively excited kink motions. 
Suppose that the large-time wave patterns in the $x=0$ plane 
    can be imaged with adequate spatial resolution. 
For definiteness, we restrict ourselves to the situation
    where the wave patterns are confined to 
    some narrow sector $S$.
Let $\partial S$ denote the outer edge of $S$.
Now suppose that only rightward wavefronts can be discerned in the immediate 
    neighborhood of $\partial S$.
It follows from Fig.~\ref{fig_3DIVP_yzplane} and Table~\ref{tab_WPsolutions} 
    that $\partial S$ can be identified as line ${\rm I}_{+}$.
Likewise, $\partial S$ is attributable to line ${\rm I}_{-}$ if both leftward
    and rightward wave fronts populate the vicinity of $\partial S$.
Regardless, the slope of $\partial S$ can then be inverted for the 
    density contrast $\rhoi/\rhoe$, whose direct measurement 
    is known to be non-trivial in the optically thin regime
    \citep[see e.g.,][and references therein]{2017ApJ...842...38X}.
Furthermore, that the wave patterns are confined to $S$ means
    that the initial exciter generates primarily 
    transverse fundamental modes. 
One therefore deduces that the spatial extent of 
    the initial exciter must be rather broad such that
    the resulting $[k_y, k_z]$ combinations lie primarily
    in the forbidden zone for the first overtone
    (see Fig.~\ref{fig_overtone_angles}).
Specifically, one deduces a lower limit
    \begin{align*}
       \Delta_{\rm min}  
      \approx \dfrac{1}{k_z^{(1/2)}} 
      = \dfrac{\sqrt{\rhoi/\rhoe-1}}{\pi/2} d 
    \end{align*}
    for this spatial extent, assuming   
    that the initial exciter is largely isotropic 
    in the $y-z$ plane. 
Recall that the excitation of higher overtones 
    demands an even narrower spatial distribution of the exciter, given that
    the radius of the forbidden zone in the $[k_y, k_z]$ plane
    scales linearly with $J$ 
    ($J=1/2, 3/2, \cdots$, see the discussion on Fig.~\ref{fig_charSols}).
A value can be further assigned to $\Delta_{\rm min}$
    in physical units with the deduced density contrast $\rhoi/\rhoe$,
    provided that the slab half-width $d$ is measurable.
Our discussions here are not meant to be exhaustive.
Rather, with the last aspect we note that the morphological features of 
    impulsively excited kink motions are potentially useful 
    for probing the initial exciters, thereby complementing
    the customary seismic practice that focuses on deducing the parameters 
    of the wave hosts. 
}

\section{Summary}
\label{sec_conc} 
Although accepted to be important,
    group velocities apparently have received little attention for three-dimensional (3D) MHD waves in solar coronal seismology.
This study offered a rather systematic examination on the behavior of group velocities
    of trapped 3D kink modes in a slab configuration, working in the framework of linear, ideal, pressureless MHD.   
Our equilibrium was taken to be structured only in one transverse direction 
    and in a piecewise constant manner.
The simplicity of the ensuing dispersion relation
    (DR, Eqs.~\eqref{eq_DR_coth} and \eqref{eq_DR_cot}) allowed substantial analytical progress, which then enabled a better understanding of the numerical solutions.
We capitalized on the computed group velocities to characterize both
    the transverse fundamental and the first transverse overtone. 
Our numerical results are placed in the context of 3D kink motions
    impulsively excited by localized perturbations, 
    the method of stationary phase (MSP) being key. 

Our findings are summarized as follows.
We came up with a three-subgroup scheme for classifying 3D kink modes on the plane spanned
    by the axial and out-of-plane wavenumbers.
The group ($\vecvgr$) and phase velocities ($\vecvph$) 
    lie on the same side of the equilibrium magnetic field ($\vec{B}_0$) 
     {for the ``$\vec{B}_0$-same-side A'' and ``$\vec{B}_0$-same-side F'' subgroups}, which are further discriminated by the proximity between $\vecvgr$ and $\vec{B}_0$.
The  {``$\vec{B}_0$-straddling'' subgroup} possesses the peculiarity that 
    $\vecvgr$ and $\vecvph$ lie astride $\vec{B}_0$, 
    a feature absent for waves in unbounded uniform media in pressureless MHD.   
This  {``$\vec{B}_0$-straddling'' subgroup} pertains to both the fundamental and its overtones.
We demonstrated how to employ the MSP to connect the distributions of
    $\vecvgr$ and $\vecvph$ in the wavenumber space to the large-time wavefront morphology
    in configurational space.
The distinction between  {``$\vec{B}_0$-straddling'' and ``$\vec{B}_0$-same-side'' modes} enabled us to categorize 
    the plane of symmetry of the equilibrium slab into different sectors.
Wavefronts directed toward $\vec{B}_0$ are confined to narrow sectors, as is the case
    for all wavefronts that can be attributed to the transverse fundamental.

Before closing, some words seem necessary to address the slab configuration
    adopted in this study.
This equilibrium is admittedly idealized.
However, we argue that slab configurations remain a useful prototype
    in solar contexts, as exemplified by its applications to wave motions observed in, say, post-flare supra-arcades~\citep{2005A&A...430L..65V}, active region arcades~\citep{2015ApJ...804L..19J,2019ApJ...880....3A}, and streamer stalks~\citep{2010ApJ...714..644C,2020ApJ...893...78D}.
Likewise, slab configurations remain to be explored theoretically, 
    with two exemplary refinements being that the equilibrium quantities
    are distributed asymmetrically \citep[e.g.,][]{2017SoPh..292...35A,2022ApJ...940..157C,2025ApJ...988...38T}
    or continuously \citep[e.g.,][]{2021SoPh..296...95Y,2022Physi...4.1359S,2025ApJ...986...40C}.
More importantly, our study can be seen as a step forward toward
    broadening the range of applicability of coronal seismology.
Aside from the customary time-series data,
    such morphological information as the directivity of propagating wavefronts
    can be of seismological use as well.

\begin{acknowledgements}
This research was supported by the 
    National Natural Science Foundation of China
    (12373055,     
     12273019,     
     12203030,     
     and 
     42230203).    
We gratefully acknowledge ISSI-BJ for supporting the international team
    ``Magnetohydrodynamic wavetrains as a tool for probing the solar corona'', and ISSI-Bern for supporting the international team 
    ``Magnetohydrodynamic Surface Waves at Earth's Magnetosphere and Beyond''.
\end{acknowledgements}

\bibliographystyle{aa}
\bibliography{Bib_up2date}

\end{document}